\documentclass[12pt]{article} 
\usepackage{amsfonts, amssymb, amsmath} 
\usepackage{fouriernc} %

\renewcommand{\bar}{\overline}

\newcommand{\Bf}{{\cal B}} %%Generic bilinear form
\newcommand{\B}{{\cal B}} %%Same again
\newcommand{\C}{{\cal C}}  %%Generic category
\newcommand{\A}{{\boldsymbol{\mathfrak A}}} %%{\frak A}} %%Generic test space

\newcommand{\G}{{\cal G}}  %%A big group; section 5...
\newcommand{\E}{{\bf E}}   %%Order-unit space arising from a probabilistic model
\renewcommand{\H}{{\bf H}}   %%Generic Hilbert space
\newcommand{\M}{{\bf M}}   %%Generic vector space
\newcommand{\p}{ P }     %%Generic projection operator 
\renewcommand{\L}{{\cal L}}
\newcommand{\V}{{\bf V}}
\newcommand{\W}{{\bf W}}

\newcommand{\Aut}{\text{Aut}}
\newcommand{\id}{\text{id}}
\newcommand{\Aff}{\text{Aff}}

\newcommand{\Hom}{\text{Hom}}
\newcommand{\Tr}{\text{Tr}}

\newcommand{\maxtensor}{\otimes_{\text{max}}}
\newcommand{\cov}{\text{cov}}

\renewcommand{\Bbb}{\mathbb} 
\renewcommand{\frak}{\mathfrak}

\begin{document}

\begin{center}{\bf Symmetry, Self-Duality, and the Jordan Structure of Quantum Mechanics}\\
%{\em Preliminary Draft, October 24 2011\\} 
\end{center}

\begin{center}
Alexander Wilce\\

%July, 2011

\end{center}

\begin{abstract} I explore several related routes to deriving the Jordan-algebraic structure of finite-dimensional quantum theory from more transparent operational and physical principles, mainly involving ideas about the symmetries of, and the correlations between, probabilistic models. The key tool is the Koecher-Vinberg Theorem, which identifies formally real Jordan algebras with finite-dimensional order-unit spaces having homogeneous, self-dual cones.\\ \end{abstract}

{\Large \bf 0 \hspace{.1in} Introduction} 

These notes  pull together some ideas for motivating the Jordan-algebraic structure of finite-dimensional quantum theory from principles having a more obvious operational or probabilistic meaning. The key tool is the Koecher-Vingerg theorem, wich lets us identify formally real Jordan algebras with finite-dimensional order-unit spaces with homogeneous, self-dual cones. The strategy is to motivate 
homogeneity and self-duality of the cone of ``effects" associated with a general probabilistic model, in terms of independently meaningful (and, ideally, plausible) principles. 

Rather than offering a single set of axioms from which this structure can cleanly be derived,  I explore in some detail the consequences of various  assumptions, mainly to do with the symmetries of a system, and with the possibility of correlating this system with a canonical ``conjugate" system. Afterwards, I observe that several 
different axiomatic packages can be extracted from the results of this study,  any of which will enforce the homogeneity and self-duality of the cone generated by a system's basic measurement outcomes.\footnote{This is in accord with a prejudice of mine, namely, that quantum theory (at any rate, its probabilistic framework) {\em does not have} a single, stark physical meaning,  but is more analogous to, say, the class of normal probability distributions, which arise in many different contexts {\em for many different reasons} --- but which can be characterized in ways that lead us to expect this ubiquity. It also reflects the conviction that the various conditions considered here, and the structures that they constrain, are of independent interest, and merit a systematic study.}

In a bit more detail, a finite dimensional {\em probabilistic model} specifies a set of basic measurements, a (compact) convex set of states --- understood as probability weights on measurement outcomes --- and a compact group of symmetries under which the both the set of measurements and the set of states are invariant. Any such model $A$ gives rise, in a canonical manner, to an order-unit space $\E(A)$, in which the positive cone is generated by the model's measurement outcomes. 
Any normalized, positive linear functional on $\E(A)$ gives rise, by restriction, to a probability weight on measurement 
outcomes. I call the model {\em state-complete} if its state space contains every such weight. 

Call a model {\em bi-symmetric} iff the group of symmetries acts transitively on pairs of distinct measurement outcomes, and on pure states. Where the state space is irreducible, bi-symmetry implies the existence of at most one $G$-invariant bilinear form on $\E(A)$ that is positive on $\E(A)_+$ and simultaneously orthogonalizes distinct measurement outcomes. Moreover, if it exists, this form is an inner product. If the model is also state-complete, it follows that the self-duality of the cone is equivalent to a condition called {\em sharpness}: every measurement outcome has probability one in a unique state. 

It remains to secure the existence of an orthogonalizing invariant, positive form on $\E(A)$. I suggest three (related) ways 
of doing so. One is to postulate the existence, for every system $A$, of a {\em conjugate} system $\bar{A}$, canonically 
isomorphic to $A$, and a bipartite non-signaling state between $A$ and its conjugate in which every measurement is perfectly, 
and uniformly, correlated with its image in $\bar{A}$. This state (analogous to the Bell state in quantum mechanics) then 
gives rise to the required bilinear form on $\E(A)$. Another approach is to require the existence of a bi-symmetric composite 
of two copies of $A$, and an invariant state in which the two component systems are independent. A third is to ask that 
all systems under consideration be representable as a set of objects in a dagger-monoidal category. 

This work builds upon the earlier papers \cite{BW, BGW, BDW, Wilce10, Wilce11b}. In particular, it echoes, but improves upon, the last of these. While I have included enough detail to make this paper reasonably self-contained, I do assume the reader has at least a glancing familiarity with the lingo of ordered vector spaces and convex cones, and more or less remembers what a Jordan algebra is. The book \cite{FK} by Faraut and Koranyi contains an excellent introduction to homogeneous self-dual cones and Jordan algebras, and includes a very accessible proof of the Koecher-Vinberg Theorem. See also \cite{Baez} for a recent survey of this material with particular reference to quantum theory. \\

\section{Order-Unit Spaces and Probabilistic Models}  

Let me begin by fixing some notation and terminology, recalling along the way some basic facts about ordered vector spaces. First, a convention: absent 
any statement to the contrary, {\em all vector spaces considered here are finite dimensional}. The dual space of a (finite-dimensional) vector space $\V$ is denoted $\V^{\ast}$; the space of linear transformations $\V \rightarrow \W$ is denoted by 
${\cal L}(\V,\W)$, with $\L(\V)$ abbreviating $\L(\V,\V)$. 
%(and understood to carry their unique locally convex topologies). 

By a {\em cone} in a real vector space $\V$, I will always mean a closed, convex, pointed, generating cone $K$ --- that is, a topologically closed convex set $K \subseteq \V$, closed under multiplication by non-negative scalars, satisfying $K \cap -K = \{0\}$, and spanning $\V$ (whence, $\V = K - K$). An {\em ordered vector space} is a real vector space 
$\V$ with a distinguished cone $K =: \E_+$ This determines a translation-invariant partial order, given by $a \leq b$ iff $b - a \in \V_+$; thus, $\V_+ = \{ a \in \V | a \geq 0\}$. The {\em standard} or {\em pointwise cone} of $\V = {\Bbb R}^{X}$ is the cone of non-negative functions. The standard cone in the space ${\cal L}(\H)$ of Hermitian 
operators on a (real or complex) Hilbert space $\H$ consists of such operators of the form $aa^{\ast}$. 

If $\V$ and $\W$ are two ordered vector spaces, a linear mapping $\phi : \V \rightarrow \W$ is {\em positive} iff 
$\phi(\V_+) \subseteq \W_+$. Note that this is a cone. If $\phi$ is a linear isomorphism and $\phi^{-1}$ is positive --- equivalently, if $\phi(\V_+) = \W_+$ --- then $\phi$ is an {\em order-isomorphism} between $\V$ and $\W$. An order-isomorphism $\V \simeq \V$ is an order {\em automorphism} of $\V$.  
%I'll denote the group of all order-automorphisms of $\V$ by 
%$\Aut(\V)$. 
The {\em dual cone} $\V^{\ast}_{+}$ is the set of positive linear functionals $f \in \V^{\ast}$. 

%If $\Omega$ is any convex set, let $\Aff(\Omega)$ denote the space of affine (that is, convex-combination preserving) %functionals $f : \Omega \rightarrow {\Bbb R}$. If $\Omega$ is compact and has finite affine dimension, then 
%$\Aff(\Omega)$ is finite dimensional (indeed, has dimension equal to the affine dimension of $\Omega$, plus $1$). Giving 
%$\Aff(\Omega)$ its obvious pointwise order, a functional $u \in \Aff(\Omega)$ 

An {\em order unit} on $\V$ is a positive functional $u \in \V^{\ast}_+$ that is {\em strictly} positive, 
i.e, $u(x) = 0$ for $x \in \V_+$ only if $x = 0$. This is equivalent (in finite dimensions, anyway), to the condition that, 
if $f \in \V^{\ast}_+$, then $f \leq nu$ for some $n \in {\Bbb N}$. More generally, an order unit {\em in} an ordered vector space $\E$ is an element $u \in \E_+$ such that, for every $a \in \E_+$, there exists $n \in {\Bbb N}$ with $a \leq nu$. An {\em order-unit space} is a pair $(\E,u)$ where $\E$ is an ordered vector space and $u \in \E_+$ is an order unit, Order unit spaces arise very naturally (as we'll see below) as {\em probabilistic models}. One defines an {\em effect} to be an element $a \in \E_+$ with $a \leq u$; a discrete {\em observable} on $\E$ is a set  $\{a_i\}$ of effects with  $\sum_{i} a_i = u$. A {\em state} on $\E$ is a positive functional $\alpha \in \E^{\ast}_+$ with $\alpha(u) = 1$, so that, for any observable, the mapping $a_i \mapsto \alpha(a_i)$ defines a probability weight on every observable $\{a_i\}$. We speak of $\alpha(a_i)$ as the probability of $a_i$ occuring when the observable $\{a_i\}$ is measured in state $\alpha$. 

As an illustration, if $\H$ is a finite-dimensional complex Hilbert space, let $\L(\H)$ denote the space of Hermitian operators on $\H$, ordered by the cone of positive operators (operators of the form $aa^{\ast}$). Then $({\cal L}(\H), \Tr)$ is an order-unit space, in which the observables are exactly the discrete ``POVMs" representing quantum observables, and the states are the linear functionals $a \mapsto \Tr(\rho a)$ corresponding to density operators $\rho$ on $\H$. 

The order unit space $\L(\H)$ has two very striking geometric properties:

%%\newpage
{\bf Definition 1 (Self-Duality and Homogeneity):} An order-unit space $\E$ is {\em self-dual} iff there exists an inner product $\langle, \rangle$ on $\E$ such that \[\E_+ = \E^{+} := \{ \ a \in \E \ | \ \langle a, b \rangle \geq 0 \ \forall b \in \E_+ \ \}.\]
An order-unit space $\E$ is {\em homogeneous} iff the group of order-automorphisms (invertible positive mappings with positive inverses) $\E \rightarrow \E$ acts transitively on the {\em interior} of $\E_+$. 

Beyond ${\cal L}(\H)$, the cone of (Jordan) squares in any formally real Jordan algebra is homogeneous and self-dual. Remarkably, this is the only example! 

{\bf Theorem (Koecher \cite{Koecher}, Vinberg \cite{Vinberg})} {\em Let $\E$ be an homogeneous, self-dual (HSD) order-unit space. Then there exists a unique formally real Jordan product on $\E$, with respect to which $\E_+$ is the cone of squares, and $u$ is the identity.} 

Both homogeneity and self-duality seem a bit more transparent to physical (or operational, or probabilistic) intuition, than does the Jordan product. So it's reasonable to try to motivate these two constraints  independently. Homogeneity seems to present the easier challenge. Indeed, if we view order-automorphisms of $\E$ as representing reversible physical processes on the corresponding system, then the homogeneity of the ``state cone" $\E^{\ast}_+$  simply requires that every interior {\em state} be reversibly transformable into any other by some physical process.
Of course, the adjective ``interior" is annoying here. In an earlier paper \cite{BGW} with Howard Barnum and Philipp Gaebler, it is shown that homogeneity also follows from the assumption that every state on $\E$ is the marginal of a bipartite ``steering" state. This condition also makes the state cone {\em weakly} self-dual, that is, isomorphic to its dual cone. However, strict self-duality requires this isomorphism to be mediated by an inner product, and this has proved trickier to motivate. 

A different approach, explored in \cite{Wilce11b}, is to derive the homogeneity and self-duality of the ``effect cone" $\E_+$ from ideas about the symmetries of systems, and the possibility of correlating two copies of a system. In order to achieve this, I made use of an ad-hoc ``minimization" axiom (which I'll review below). Here, I aim to do better, and, in particular, 
to avoid this assumption.   \\

%[Comment: homogeneity has a certain immediate physical appeal. So does a weak form of self-duality (explain). 
%And, in the context of a WSD model, we gain some further insight into homogeneity: as I'll discuss further below, 
%a positive linear mapping $\V^{\ast} \rightarrow \V$ can be regarded as a bipartite state, that is, a state (possibly %entangled) on a pair of copies of $\V$. [Well, w/ LT..]; hence, homog. can be reformulated as the 
%%statement that, for any effect $a$ and state $\alpha$ in the interior of $\V^{\ast}, \V$ resp., there is 
%a state $\omega$ such that, conditioning on the occurence of $a$ on one copy, we obtain state $\alpha$ on the other. 
%Equivalently (reversible...) 

%Steering...   

%... ETC. Now the point that real self-duality requires an inner product!]

%%\newpage
{\bf 1.1 Test spaces and probabilistic models} 

For my purposes, the abstract order-unit spaces dealt with above are a little {\em too} abstract. 
% or, to put it another way, they have too little structure. 
%I will work instead with more concrete objects I call {\em probabilistic models}, to which 
%order unit spaces are attached as invariants. 

%Let's begin with a bare-bones operational framework. 

{\bf Definition 2 (Test spaces):} A {\em test space} is a pair $(X,\A)$ where $X$ is a set of {\em outcomes} and $\A$ is a covering of $X$ by non-empty sets called {\em tests}, interpreted as the sets of mutually exclusive outcomes associated with various experiments. A {\em probablity weight} on $(X,\A)$ is a function $\alpha : X \rightarrow [0,1]$ with $\sum_{x \in E} \alpha(x) = 1$ for every $E \in \A$.  I'll write $\Omega(X,\A)$ for the convex set of all probability weights on $(X,\A)$.\footnote{Mathematically, a test space is just a hypergraph. The terminology is meant to enforce a particular interpretation. Test spaces --- originally termed ``manuals" --- were the basis for a generalized probability theory (and an associated ``empirical logic") developed in the 1970s and 80s by C. H. Randall and D. J. Foulis and their students. See \cite{Wilce10} for a survey.  It is important to understand that $\A$ is not necessarily intended as the complete catalogue of {\em all} possible measurements on a given system, but only some set of measurements sufficient to capture the system's states, 
which we have singled out for some reason (perhaps one of tradition, or of exegetical efficiency).}

The {\em rank} of a test space $(X,\A)$ is the least upper bound of $|E|$ where $E \in \A$. For purposes of this note, {\em all test spaces have finite rank.} In particular, all tests are finte sets. It follows easily that the set 
$\Omega(X,\A)$ of all probability weights on $(X,\A)$ is a closed, and hence, compact, subset of $[0,1]^{X}$. 

{\em Notation:} Anticipating later results, I'll write $x \perp y$ to mean that outcomes $x, y \in X$ are {\em distinguishable} by means of a test in $\A$ --- that is, that $x \not = y$ and there exists some $E \in \A$ with $x, y \in E$. Note that, at present, there is no linear structure in view, let alone an inner product, so the notation 
is only suggestive. 

{\bf Definition 3 (Symmetry):} By a {\em symmetry} of a test space $(X,\A)$, I mean a bijection $g : X \rightarrow X$ that permutes elements of $\A$. 

Notice that if $g$ is a symmetry, then for all $x, y \in X$, $x \perp y$ iff $gx \perp gy$.  An action of a group $G$ on $(X,\A)$ is an action by symmetries, and a test space equipped with such an action is a {\em $G$-test space}. I'll write $\Aut(X,\A)$ for the group of all symmetries of $(X,\A)$.  In the cases that will interest 
us, this will always be isomorphic to a compact subgroup of $GL(d)$ for a sufficiently large finite dimension $d$. 

In constructing a model for a probabilistic system, we may want to privilege not only the ``observables" represented by the tests $E \in \A$, but also certain states and certain symmetries,  This suggests the following

{\bf Definition 4 (Probabilistic Models):} A {\em probabilistic model} --- or, for purposes of this note, just a {\em model} --- is a structure $(X,\A,\Omega,G)$, where $(X,\A)$ is a (finite-rank)  test space, $\Omega$ is a separating, pointwise-closed (hence, pointwise compact) convex set of probability weights on $(X,\A)$, and $G$ is a compact group of symmetries of $(X,\A)$ leaving $\Omega$ invariant. 

I'll call $\Omega$ the {\em state space} of the model; probability weights $\alpha \in \Omega$ are {\em states}. Where $\Omega$ has finite affine dimension, I'll say that  the model is finite-dimensional. {\em All models considered in this paper are finite-dimensional in this sense.}
%\footnote{What I'm calling a model is a refinement of Mackey's original set-up for generalized probability, consisting of a family of ``questions" (essentially, two-outcome tests) and a family of probability weights thereon. The main novelty here is the specification of the group $G$.}  
In the interest of sanity, I'll hereafter denote models by Roman capital letters $A, B, ...$, writing (for instance) $A = (X,\A,\Omega, G)$. It will often be convenient to label the components with the name of the model, as, e.g., $(X(A), \A(A), \Omega(A), G(A))$. I will use the terms ``model" and ``system" interchangeably.

It's time to look at some examples.

{\bf Example 1: Classical models}  Let $E$ be a single, classical outcome-set (say, for a coin-flip, or rolling a die). Let $X = E$, $\A = \{E\}$, and $G \leq S(E)$ be any group you like of permutations 
of $E$. Let $\Gamma$ be any separating, permutation-invariant set of probability weights on $E$, and let $\Omega$ be the closed convex hull of $\Gamma$. Alternatively, choose any separting closed convex set $\Omega$ of probability weights, and let $G$ be the group of permutations leaving $\Omega$ invariant. 

{\bf Example 2: Quantum Models} Let $\H$ be an $n$-dimensional complex Hilbert space. The corresponding 
{\em quantum model} is $A(\H) := (X(\H),\A(\H),\Omega(\H),U(\H))$, where 
\begin{itemize}
\item[$\bullet$] $X(\H)$ is the set of rank-one projection operators 
on $\H$, 
\item[$\bullet$] $\A(\H)$ is the set of maximal pairwise orthogonal sets of such projections, 
\item[$\bullet$]  $\Omega(\H)$ is the set of states of the form  $x \mapsto \Tr(\rho x)$,  
 $\rho$ a density operator on $\H$\footnote{Gleason's Theorem tells us that $\Omega(\H) = \Omega(X(\H),\A(\H))$ for 
$\dim(\H) > 2$; for $\dim(\H) = 2$, the density matrices need to be put in by hand.} and 
\item[$\bullet$] $U(\H)$ is the group of unitary operators on $\H$, acting on $X$ by conjugation.
\end{itemize} 

{\bf Example 3: The Square Bit} For a much different, and much simpler, example, consider a test space $(X,\A)$ consisting of two disjoint, two-outcome tests --- say, $X = \{a,a',b,b'\}$ and $\A = \{\{a,a'\}, \{b,b'\}\}$. Then the space $\Omega$ of all probability weights on $(X,\A)$ is affinely isomorphic to the unit square in ${\Bbb R}^{2}$. The {\em square bit} is the model $(X,\A,\Omega,G)$ where $G$ is the dihedral group acting on $\Omega$ in the obvious way, and dually on $(X,\A)$. 

{\bf Example 4: Jordan Models} Let $\E$ be a formally real Jordan algebra, and let $X$ denote the set of primitive idempotents in $\E$. A {\em Jordan frame} is a pairwise orthogonal set of idempotents 
summing to the order unit. Letting $\Omega$ denote the set of states on $\E$ and $G$, the set of Jordan automorphisms of $\E$, we have a {\em Jordan model} $(X,\A,\Omega,G)$. \\

{\bf 1.2 Models Linearized} 

Let $A$ be a  model. Every outcome $x \in X(A)$ determines an affine functional $\hat{x} : \Omega(A) \rightarrow {\Bbb R}$ by evaluation: $\hat{x}(\alpha) = \alpha(x)$. Letting $\Aff(\Omega(A))$ denote the space of all real-valued affine functionals on $\Omega(A)$, we have then a natural --- and, clearly, $G$-equivariant --- mapping 
$X(A) \rightarrow \Aff(\Omega(A))$. It is largely harmless to assume that this is injective, i.e., that $\Omega(A)$ 
separates points of $X$. (If not, replace $(X(A),\A(A))$ by the obvious quotient structure.) From now on, I assume 
this is the case; that is, I make it a {\em standing assumption} that {\em all probabilistic models have separating sets of states.} 

In view of this, it is convenient to identify $x \in X(A)$ with the corresponding functional, so that $X \subseteq \Aff(\Omega(A))$. I also assume, from this point on, that {\em all models are finite dimensional}, in the sense that $\Omega(A)$ has finite affine dimension. It follows that $\Aff(\Omega)$ is a finite-dimensional real vector space. Let $\E = \E(A)$ denote the span of $X$ in $\Aff(\Omega)$, ordered by the 
cone consisting of linear combinations of outcomes having non-negative coefficients:
\[\E_{+} = \{ \ \sum_{i} t_i x_i \ | x_i \in X, \ t_i \geq 0 \ \}.\]
Note that this may be smaller than the cone $\{ \ a \in \E \ | \ a(\alpha) \geq 0 \ \forall \alpha \in \Omega \ \}$ inherited from $\Aff(\Omega)_{+}$, 
and that, unlike the latter, it depends on the choice of $X$. Notice, too, that the action of $G$ on $X$ extends uniquely to a linear action on $\E$, given by $(ga)(\alpha) = a(\alpha \circ g)$ for all $a \in \E$ and all $\alpha \in \Omega$, and that $\E_+$ is stable under this action. Finally, observe that, for every $E \in \A$, $\sum_{x \in E} x = u$, where $u$ is the unit functional 
$u(\alpha) \equiv 1$ for all $\alpha \in \Omega$. This last serves as an order-unit for $\E$. 

I'll call the order-unit space $(\E(A),u)$ the {\em linear hull} of the model $A$. Notice that every test $E \in \A(A)$ can now be regarded 
as a discrete observable on $\E(A)$. Thus, we can, to a large extent, regard a probabilistic model as an order-unit space equipped 
with a distinguished collection of observables (sufficient to separate points), invariant under a distinguished 
compact group of order-automorphisms, {\em and} with a distinguished convex set $\Omega$ of states (of which, more in a moment). 
%However, I regard the test space $(X,\A)$ as being at least as fundamental as the state space $\Omega$. (Certainly, one can imagine situations 
%in which one might wish to hold the former fixed while varying the latter.) Accordingly, I'll treat the space $\E(A)$ as a useful invariant of the model $A$, keeping the latter in the foreground. 
% prefer, 
%however, to keep the operational infrastructure of the probabilistic model in view. 
%In order to emphasize that the state space will, 
%in general, be smaller than 

%Notice that any set ${\cal E} \subseteq \E(A)_+$ HERE: EFFECTS, OBSERVABLES...] 

{\bf Examples:} In the case of a quantum model $A = A(\H)$, $\E(A)$ can be identified with the order-unit space ${\cal L}(\H)$ of Hermitian operators on $\H$, ordered by the usual cone, with  $u$ the identity operator.  In the case of the square bit, $\E(A)$ is isomorphic to ${\Bbb R}^{3}$, equipped with a cone having a square cross-section. In the case of a Jordan model, $\E(A)$ is canonically isomorphic, as an order-unit space, to the given Jordan algebra.\\

%%\newpage
{\bf 1.3 Sharpness and State-Completeness} %[Clean up...] %[HERE $\E_+$ is the positive span of $X$ ...] 

If $(\E,u)$ is any order-unit space, a {\em state} on $\E$ is a positive linear functional $\rho : \E \rightarrow {\Bbb R}$ that is normalized so that $\rho(u) = 1$. 
If $A = (X,\A,\Omega,G)$ is a model, with linear hull $\E(A)$, then any state $\alpha \in \Omega$ defines a state on $\E(A)$, just by evaluation: $\alpha(a) := a(\alpha)$. 
Conversely, a state $\rho$ on $\E$ defines a state on the test space $(X,\A)$ by restriction. In general, however, this {\em will not} lie in the designated state space $\Omega(A)$ of the model. 

Let $\widehat{\Omega(A)}$ denote the set of all states on $(X,\A)$ arising from states on $\E(A)$. Obviously, $\Omega \subseteq \widehat{\Omega}$. We may regard $\widehat{\Omega}$ as the set of probability weights that are consistent with all of the linear relations among outcomes that are satisfied by the given state space $\Omega(A)$. Evidently, the assignment $\Omega \mapsto \widehat{\Omega}$ is a closure on the poset of closed convex subsets of the (full) state space of $(X,\A)$. Let's agree to call a model {\em state-complete} iff $\Omega = \widehat{\Omega}$.  In this case, $\E_+$ coincides with the cone $\E \cap \Aff(\Omega(A))_+$, i.e, every element of $\E_+$ taking positive values on $\Omega$ belongs to $\E_+$. (This last condition is called {\em saturation} in \cite{BW}). 

{\bf Example 5:} For an example of a non-state complete model, let $E = \{x,y\}$ be a single two-outcome classical outcome-set, and consider the probability weights $p_1, p_2$ given by 
$p_1(x) = .6, p_1(y) = .4$, and $p_2(x) = .4, p_2(y) = .6$. The set $\{p_1, p_2\}$ is invariant under the obvious action of $G = S_2 = S(E)$, and separates $x$ and $y$. Let $A = (E,\{E\}, \Omega, S_2)$ where $\Omega$ is the closed convex hull of $p_1$ and $p_2$, i.e., $\Omega = \{ t p_1 + (1 - t)p_2 | 0 \leq t \leq 1.\}$. Then $\E(A) \simeq {\Bbb R}^{2}$ with $\E_+$ the first quadrant. The full state space $\hat{\Omega}$ consists of all probability weights on $E$, and is thus considerably larger than $\Omega$.\footnote{Thanks to Jon Barrett for pointing out this sort of simple example.}   

All of the non-classical models discussed above are state-complete. State-completeness is a pretty reasonable condition to impose on a probabilistic model, at least in a finite-dimensional setting, and it will figure as a crucial hypothesis in many of the results below. Nevertheless, in order to keep clearly in view what does and what does not depend on it, I make {\em no standing assumption} of state-completeness.  To help in keeping this in mind, I'll use the notation $\V(A)$ for the space $\E(A)^{\ast}$, ordered not by the natural dual cone, but by 
the cone $\V_+(A)$ generated by the designated state space $\Omega(A)$. State-completeness amounts to the condition that $\V(A) = \E(A)^{\ast}$ (in which case, we also have 
$\E(A) \simeq \V(A)^{\ast}$.

%HERE: SC and self-duality. 

%The following simple observation will be useful in the sequel. If $\B$ is any bilnear form (not necessarily and inner product) %on $\E$, let $\E^{+} = \{ a \in \E ~|~ \B(a,b) \geq 0 \ \forall b \in \E_+\}$. I'll say that $\E$ is {\em self-dual with %respect to $\B$} iff $\E^{+} = \E_+$, and that $\B$ is {\em positive} iff $\B(a,b) \geq 0$ for all $a, b \in \E_+$ --- or, what %is the same thing, if $\E_+ \subseteq \E^{+}$. 

%{\bf Lemma 0:} {\em Let $A$ be a probabilistic model, and let $\tau : \E(A) \rightarrow \V(A)$ be an order-isomorphism. 
%Let $\B(a,b) = \tau(a)(b)$. Then $\E$ is self-dual with respect to $\B$ iff $A$ is state-complete.}

%{\em Remark:} Writing $\Sigma(\E)$ for the state space of an order-unit space $\E$, the mapping $\Omega(A) \mapsto \Sigma(\E(A))$ is a closure on the set of closed convex subsets of $\Omega(X,\A)$. %    Replacing $\Omega$ with $\Sigma(\E(A))$, we obtain a new, state-complete model, having the same linear hull.  
Another condition that will play a significant role in what follows is {\em sharpness}:

{\bf Definition 5 (Sharpness):} A model $A = (X,\A,\Omega,G)$ is {\em sharp} iff, for every $x \in X$, there exists a unique 
state $\delta_x \in \Omega$ with $\delta_x(x) = 1$. 

In the earlier papers \cite{Wilce10, Wilce11b}, I called a model sharp iff, for every outcome $x \in X(A)$, there exists a unique 
state $\alpha \in \E^{\ast}(A)$ with $\alpha(x) = 1$. If $A$ is state-complete (as was tacitly assumed in \cite{Wilce11a}), this coincides with the oresent notion. Sharpness (in one form or another) has a long history in the 
quantum-logical literature. In particular, it played a central role in Gunson's axiomatics for quantum theory \cite{Gunson}. A stronger form of sharpness, in which it is also required that each pure state render certain a unique outcome, is used by Hardy in \cite{Hardy2}. \\

%(A stronger condition would be to require not only sharpness, but also the principle that, for 
%every pure state $\alpha$, there is a unique outcome with $\alpha(x) = 1$.)    \\ 

{\bf 1.4 Morphisms of Models} 

At several points I'm going to need to treat models categorically. Thre are various notions of morphism one might use, but the one that makes the most sense in the current context seems to be the following.   

%If $\phi : X \rightarrow Y$, write $\phi^{\ast} : {\Bbb R}^{Y} \rightarrow {\Bbb R}^{X}$ given by $\phi^{\ast}(f) = f \circ \phi$. 

{\bf Definition 6 (Morphisms):} A {\em morphism} from a model $A$ to a model $B$ is a pair $(\phi,\psi)$, where  
\begin{itemize} 
\item[(i)] $\phi$ is a mapping $X(A) \rightarrow X(B)$, pushing tests of $A$ forward to tests of $B$, and pulling states of $B$ back to states on $A$ --- that is, 
\[\phi(\A(A)) \subseteq \A(B) \ \text{and} \ \phi^{\ast}(\Omega(B)) \subseteq \Omega(A)\]
(where $\phi^{\ast} (\beta) = \beta \circ \phi$.) 
\item[(ii)] $\psi \in \Hom(G(A),G(B))$; 
\item[(iii)] $\phi(gx) = \psi(g)\phi(x)$ for all $x \in X(A), g \in G(B)$. 
\end{itemize} 

In practice, it will be convenient to regard $\psi$ as defining an action of $G(A)$ on $X(B)$, writing $gy$ for $\psi(g)y$ for $g \in G(A)$ and $y \in X(B)$. When I wish to suppress explicit mention of 
$\psi$ in this way, I'll simply write $\phi$ for the pair $(\phi,\psi)$.  From this point of view, (iii) says that $\phi$ is equivariant. (Note, though, that the given action of $G(A)$ on $X(B)$ must be through elements of $G(B)$.) 

An {\em isomorphism} of models is an invertible morphism; equivalently, a bijective mapping $\phi : X(A) \rightarrow X(B)$, equivariant with respect to an action of $G(A)$ on $X(B)$ (by members of $G(B)$), taking $\A(A)$ bijectively onto $\A(B)$, and inducing an affine isomorphism $\phi^{\ast} : \Omega(B) \rightarrow \Omega(A)$. In particular, every symmetry of a model $A$ is a 
morphism from $A$ to itself. 

A morphism $\phi : A \rightarrow B$ lifts naturally to a positive linear mapping between the corresponding linear hulls. To spell this out, notice that the affine mapping $\phi^{\ast} : \Omega(B) \rightarrow \Omega(B)$ guaranteed by condition (i) of the definition, induces a linear map $\phi^{\ast \ast} : \Aff(\Omega(A)) \rightarrow \Aff(\Omega(B))$, given by $(\phi^{\ast \ast} a)(\beta) = a(\phi^{\ast}(\beta)) = a ( \beta \circ \phi)$. Identifying $x \in X(A)$ with the corresponding vector $x \in \E(A) \leq \Aff(\Omega(A))$, and similarly taking $\phi(x) \in \E(B) \leq \Aff(\Omega(B))$, it follows that 
$\phi^{\ast \ast}(x) = \phi(x)$. Thus, $\phi^{\ast \ast}$ restricts to a linear mapping $\phi : \E(A) \rightarrow \E(B)$ extending $\phi : X(A) \rightarrow X(B)$. Since this takes outcomes to outcomes, it sends $\E_+(A)$ into $\E_+(B)$, that is, $\phi$ is positive. Note, too, that if $E \in \A(A)$ and $F = \phi(E) \in \A(B)$, we have 
$\phi(u_{A}) = \sum_{x \in E} \phi(x) = \sum_{y \in F} y = u_{B}$. Thus, we can regard $A \mapsto \E(A)$ as the object part of a functor from probabilistic models and morphisms, to order-unit spaces and positive, unit-preserving linear maps. This observation will be put to use in due course.\\

\section{Bi-Symmetric Models} 

I now wish to impose some constraints on the models under consideration. This section spells out some consequences of a package of symmetry assumptions which, taken together, assert that (i) all pure states, all outcomes, and all tests tests look the same, and (ii) individual tests have no (or little) internal structure, in the sense that the outcomes of any test can be permuted more or less freely by symmetries of the model, keeping the test fixed. 

{\bf Definition 7a (Full symmetry):} A test space $(X,\A)$ is {\em fully symmetric} under the action of a group $G$ 
iff (i) every test $E \in \A$ has the same cardinality, and (ii) every bijection $f : E \rightarrow F$, $E, F \in \A$, is implemented by some element $g \in G$, i.e., $gx = f(x)$ for every $x \in E$. 

See \cite{Wilce05, Wilce11a, Wilce11b} for more on this notion. Full symmetry entails that $G$ act transitively on both $\A$ and $X$. 

{\bf Example 6:} Let $\E$ be a formally real Jordan algebra, and let $X$ denote the set of primitive (that is, atomic) idempotents in $\E$. Let $\A$ be the collection of all finite subsets of $X$ summing to the unit element of $\E$, 
and let $\Omega$ be the set of all $\rho \in \E_+$ with $\langle \rho, u \rangle = 1$ where $\langle \ , \ \rangle$ 
is the canonical %[ref?] 
inner product on $\E$. Finally, let $G$ be the group of all Jordan automorphisms of $\E$. 
Then $(X,\A,\Omega,G)$ is a probabilistic model. Moreover, it is fully symmetric (\cite{FK}, Theorem IV.2.5). %Cite Theorem... 

A weaker condition than full symmetry, still sufficient for most of what follows, is that $G$ act transitively on $\A$ and on the set of orthogonal (that is, distinguishable) {\em pairs} of outcomes: 

{\bf Definition 7b ($2$-Symmetry):} $(X,\A)$ is {\em $2$-symmetric} under the action of $G$ iff (i) $G$ acts transitively on $\A$, and (ii) $G$ acts transitively on pairs of distinguishable measurement outcomes, 
that is, for all outcomes $x, y, u, v \in X$ with $x \perp y$ and $u \perp v$, then there exists some $g \in G$ such that $gx = u, gy = v$. Note that any fully-symmetric test space is also $2$-transitive.

In the context, not of a test space, but of a probabilistic model, I am also going to ask that $G$ act 
transitively on the set of pure states. Thus, 

{\bf Definition 8 (Bi-symmetry):} A model $(X,\A,\Omega,G)$ is {\em fully bi-symmetric}, respectively {\em bi-symmetric}, iff (i) $G$ acts fully symmetrically, resp., $2$-symmetrically, on $(X,\A)$, and (ii) $G$ acts transitively on extreme points of $\Omega$. 

%[Do we rather want to talk about cyclic representations?]

Bi-symmetric models can readily be constructed ``by hand", as follows \cite{Wilce11a}. Suppose $E$ is a set, thought of as the outcome-set of a ``standard test", and suppose $H$ is a group acting $2$-transitively on $E$.  
Let $G$ be any group with $G \geq H$, and let $K \leq G$ be any subgroup of $G$ with $K \cap H = H_{x_o}$, the stabilizer in $H$ of some point $x_o \in E$. Set $X = G/K$, and embed $E$ in $X$ via $hx_o \mapsto hK$, where 
$h \in H$. (The condition that $K \cap H = H_{x_o}$ guarantees that this is well-defined). Let $\A$ be 
the orbit of $E$ in ${\cal P}(X)$ under $G$, that is, 
${\frak A} = \{ gE | g \in G\}$. Then $G$ acts $2$-symmetrically on $(X,\A)$. 
Now choose any $\delta_o \in \Omega(X,\A)$, and set $\Omega$ be the closed convex hull of $G \delta_o$. See \cite{Wilce05, Wilce11a} for more on this construction. The possibility of freely construcing bi-symmetric models in this way means, on the one hand, that
bi-symmetry is a reasonably benign assumption, but also that it is not a very constraining one. 

{\em Remark:} Individually, state-transitivity and state-completeness are very reasonable axioms: the former asks that we construct our state space in a natural way (as just outlined); the latter asks that we enlarge our state space, if necessary, in an equally natural way. However, there is a tension between these reasonable requirements, in that enlarging the state space to secure state-completeness may spoil state-transitivity. We can only rarely satisfy both 
conditions at once.\footnote{This should not be too dismaying: a set of axioms {\em must} be in some tension with one another if they are to single out a narrow class of models.}\\
%(where $G$ acts in the obvious way on ${\Bbb R}^{X}$). 

{\bf 2.1 SPIN forms} 

Until further notice, $A = (X,\A,\Omega,G)$ is a bi-symmetric model of rank $n$, 
and $\E = \E(A)$. %with linear hull $\E$.  

{\bf Definition 9 (SPIN forms):} Let $\Bf : \E \times \E \rightarrow {\Bbb R}$ be a bilinear form. I will say that $\Bf$ is {\em positive} iff 
$\Bf(a,b) \geq 0$ for all $a, b \in \E_+$, {\em normalized} iff $B(u,u) = 1$, and {\em invariant} iff 
$\Bf(ga,gb) = B(a,b)$ for all $g \in G$. I'll call a {\em symmetric} positive, invariant, normalized bilinear bilinar form on $\E$ a {\em SPIN form} for short.\footnote{Of course, this is an absolutely dreadful choice of terminology; but I can't seem to think of anything better at the moment. Suggestions?} . 

There is a more or less canonical example, namely, the inner product 
\[\langle a, b \rangle_G := \int_{G} a(g\delta_o)b(g\delta_o) dg\]
where $\delta_o$ is any pure state (that is, extreme point) in $\Omega$ and the integration is with respect to normalized Haar measure. Owing to the transitivity of $G$ on the set of pure states, 
this is independent of the choice of $\delta_{o}$ %[CHECK]. 
 
We can also define a {\em degenerate} SPIN form $\B_o$, defined by 
\[\B_o(a,b) = \langle a, u \rangle_G \langle b, u \rangle_G\]
for all $a, b \in \E$. This turns out to be independent of the choice of the SPIN inner product (indeed, of the SPIN form) appearing on the right. %$\langle ~,~ \rangle$. 
%about which I'll have more to say below. %[NOTE: Do we really just need $G\delta_o$ to be separating? ]

%The following identities and inequalities follow easily from the definition. 
Any SPIN form on $\E$ is associated with two non-negative real constants:
\begin{itemize} 
\item[(1)] $r^2 := \B(x,x)$ for all $x \in X$, and 
\item[(2)] $c := \B(x,y)$ for all $x, y \in X$ with $x \perp y$. 
\end{itemize} 
Call these the {\em parameters} of $\B$ (though, as we'll now see, they are not independent of one another).

{\bf Lemma 1:} {\em Let $\B$ be a symmetric, positive, invariant, normalized bilinear form on $\E$. 
Then the parameters $r$ and $c$ satisfy 
\begin{itemize} 
%\item[(a)] $B(x,x) := r^{2} \geq 0$ for all $x \in X$; 
%\item[(b)] $B(x,y) := c$ is a constant independent of $x \perp y$ in $X$;
\item[(a)] $\B(x,u) = 1/n$ for all $x \in X$; 
\item[(b)] $r^2 + (n-1)c = 1/n$
\item[(c)] $r^{2} \leq 1/n$. 
\item[(d)] Let $m$ and $M$ denote, respectively, the minimum and maximum values of $\B$ on $X \times X$. Then 
$m \leq 1/n^2 \leq M$. 
\item[(e)] If $\B$ is positive-semidefinite, $r^{2} \geq 1/n^2 \geq c$. 
\item[(f)] If $\B$ is an inner product and $r^2 = 1/n^2$, then $\E$ is one-dimensional. 
%B$ is the degerate 
%form $B_o$.  
%If $B$ is an inner product, we have equality only if $\dim\E = 1$. 
%\item[(e)] Let $m$ denote the minimum value of $B(x,y)$ on $X \times X$. Then $m \leq 1/n^2$. 
%\item[(g)] If $\B$ is positive semi-definite, then $c \leq 1/n^2$.
\end{itemize} }

{\em Proof:} 
%(a) and (b) are straightforward consequences of $2$-symmetry, (a) following from the tansitivity of $G$ on $X$, and %(b) from the transitivity of $G$ on orthogonal pairs.  
For (a), note that 
$\B(x,u)$ is a constant, again by transitivity of $G$ on $X$, whence,  $n\B(x,u) = \sum_{x \in E} \B(x,u) = \B(u,u) = 1$.  For (b) note that if $x \in E \in \A$, we have $\B(x,u) = \sum_{y \in E} B(x,y) = \B(x,x) + \sum_{y \in E \setminus \{x\}} \B(x,y) = r^2 + (n-1)c$, which gives the desired inequality. Since $c_{\B} \geq 0$, this also yields (c), as $r^2 \leq r^2 + (n-1)c_B$. 
For (d), note that for any $E \in \A$, we have 
$n^2 m \leq \sum_{x, y \in E} \B(x,y) = \B(u,u) = 1$; similarly, $n^2M \geq \B(u,u) = 1$.

For (e), observe that if $\B$ is positive semi-definite, then $\|v\| := \sqrt{\B(v,v)}$ is a semi-norm, with $\|x\| = r$ for every $x \in X$. Hence, by the triangle inequality, we have  
\[1 = \|u\| = \|\sum_{x \in E} x\| \leq \sum_{x \in E} \|x\| = n r,\]
whence, $r^{2} \geq 1/n^2$. It now follows from (b) that 
\[(n-1)c = 1/n - r^2 \leq 1/n - 1/n^2 = \frac{n-1}{n^2},\]
giving us $c \leq 1/n^2$.  Finallly, for (f), suppose $\B$ is an inner product. If $r^2 = 1/n$, then it follows from (b), as above, that $(n-1)c = \frac{n-1}{n^2}$, whence, $c = 1/n^2$ as well. 
Hence, for any $x \perp y$ in $X$, we have 
\[\frac{1}{n^2} = c = \B(x,y) = \|x\|\|y\|\cos \theta = r^2 \cos \theta = \frac{1}{n} \cos \theta,\]
whence, the angle $\theta$ between $x$ and $y$ is $0$, i.e., $x = y$. It follows that $(X,\A)$ has rank $n = 1$, whence, $\dim \E = 1$.  
%[CAUTION HERE!! This is only correct for an inner product. Note that the trivial form satisfies $r^2 = = 1/n$. ... So: the %inner product on the quotient space satisfies this (??) - so, must be 1-d. 
%IDEA: there's only one form like this, the trivial one.]

Finally, if $\B$ is positive semi-definite, then by (b) and (c), we also have 
\[(n-1)c \leq 1/n - r^2 \leq 1/n - 1/n^2 = \frac{n-1}{n^2},\]
giving us (g). $\Box$ 
%\footnote{In fact, the only part of full symmetry we use for Lemma 1 is what we might call $2$-symmetry: 
%for any pairs $x \perp y$ and $u \perp v$ in $X \times X$, there exists some $g \in G$ with $gx = u, gy = v$. %Indeed, almost every result that follows depends only on this weaker assumption.} Note that we use only the full %symmetry of $G$'s action on $(X,\A)$; the second condition, that $G$ act transitivity on pure states, is not used. %$\Box$

%{\bf Corollary 1:} {\em With notation and hypotheses as above, let $r = 1/n^2$. Then $n = 1$.} 

%{\em Proof:} ... [Not needed for irred. case, where we have an inner product...]

%{\em Remark:} Everything here goes through just as stated with an arbitrary representation $(E,\pi)$ in place 
%of the linear hull $\E$. The constants $r$ and $c$ defined above, I'll call the {\em parameters} 
%of the representation, or, where the representation is the canonical one, the parameters of the model. 
%Let $B$ denote any SPIN form, and let $B_{o}(a,b) = B(a,u)B(u,b)$. Then $B_o$ is a SPIN form with $B(x,y) = 1/n^2$ %for all $x, y \in X$, and is uniquely defined by this property, and independent of $B$. I'll call $B_o$ the {\em %degenerate} or {\em trivial} SPIN form on $\E$.   

{\bf Corollary 1:} {\em If $\B_1$ and $\B_2$ are SPIN forms on $\E$, then 
for all $a \in \E$, $\B_1(a,u) = \B_2(a,u)$.} 

{\em Proof:} If $x \in X$, then $\B_1(x,u) = \B_2(x, u) = 1/n$, by Lemma 1. Since $X$ spans $\E$, the result follows 
from the bilinearity of $\B_1$  and $\B_2$. $\Box$ 

A consequence is that the degenerate form $\B_o(a,b) := \B(a,u)\B(b,u)$ is independent of the choice of $\B$. Henceforth, 
I refer to this SPIN form as the {\em uniform} SPIN form. \\
%This gives 
%us a complement to part () of Lemma 1: [THE FOLLOWING HOLDS FOR IRREPS]

%{\bf Corollary 2:} {\em Let $A$ be irreducible. If $B$ is a SPIN form on $\E(A)$ satisfying $r^2 = c = 1/n^2$, then $B = B_o$.}

%{\em Proof:} 
%\\

%(Handle $1/n^2$ case separately?)

%From now on, let's write SPIN for the phrase {\em symmetric, positive, invariant, normalized}.\footnote{I realize that this is %probably a very poor choice of terminology. Suggestions of alternatives %would be welcome.}
%If $B$ is a SPIN bilinear form, write $r_B$ and $c_B$ for the corresponding parameters. 
%It is natural to require this to belong to $\Omega$. %[WORK ON %THIS!!]

%%\newpage
{\bf 2.2 Minimizing and Orthogonalizing Forms} 

If $\B$ is a SPIN bilinear form on $\E$ and $x \in X$, we can define a probability weight $\alpha_x$ on $(X,\A)$ 
by $\alpha_x (y) := n\B(x,y)$ for all $y \in X$. There is, however, no guarantee that this state will belong to the designated state space $\Omega$. 

Since $G$ is compact, and acts continuously on $\E$, its orbits are also compact. In particular, $X$ is compact. It follows 
that every bilinear form --- in particular, every SPIN bilinear form --- achieves a maximimum and a minimum value on $X \times X$. 

{\bf Definition 10 (Minimizing and Orthogonalizing SPIN forms):} A SPIN bilinear form $\B$ on $\E$ is {\em minimizing} iff $\B(x,y)$ achieves its minimum value on $X \times X$ 
at a point $(x,y)$ with $x \perp y$.  $\B$ is {\em orthogonalizing} iff $c_{\B} = 0$, i.e., $\B(x,y) = 0$ for all $x, y \in X$ with $x \perp y$.  

Clearly, orthogonalizing implies minimizing.  In the language of this paper, Proposition 1 of \cite{Wilce11b} asserts that {\em if $A$ is bi-symmetric, sharp, state-complete model, and $\E(A)$ admits a minimizing form, then $\E(A)_+$ 
is self-dual.} It is also shown that, under these assumptions, $\E(A)$ has an orthogonalizing form. One of the main goals of the present paper is to find sufficient conditions for such a minimizing form to exist.

%In [W11], the existence of a minmizing form is postulated. Here,
%the derivation of a self-dual representation for $(X,\A,\Omega,G)$. 
%In [W11], the existence of a minimizing form was taken as a postulate; 
%from this, the existence of an orthogonalizing representation was derived.  
%As shown below, in the presence of a dagger-monoidal structure, the existence of an orthogonalizing form is automatic. [Question: Does existence of min. therefore imply existence of orthog?] 

The existence of an orthogonalizing form has many consequences. For one thing, if $\B$ is orthogonalizing, then 
for every $x \in X$, the probability weight $\delta_x := n \B(x, \cdot)$ assigns probability $1$ to $x$ and 
$0$ to any outcome $y \perp x$.  

%[The following condense, omit, or move...]
If $\B$ is a SPIN form on $\E$, let's agree to write $\E^{+}$ for the set of vectors $a \in \E$ such that $\B(a,b) \geq 0$ for all $b \in \E_+$  (even if $\B$ is not an inner product).  The positivity of $\B$ guarantees that $\E_+ \subseteq \E^{+}$. 
If $\E^{+} = \E_{+}$, I'll say that $\E$ is {\em self-dual with respect to} $\B$. The following is essentially proposition 1 from \cite{Wilce11b}, but formulated more generally, 
for SPIN forms rather than SPIN inner products:

{\bf Lemma 2:} {\em Let $A$ be a sharp, state-complete, bi-symmetric model. If $\B$ is a non-degenerate orthogonalizing 
SPIN form on $\E(A)$, then $\E(A)$ is self-dual with respect to $\B$.} 
 
{\em Proof:} We have $\E_{+} \subseteq \E^{+}$ in any case. Let $x \in X$. Since $A$ is state-complete, the probability 
weight $\delta_x := n\B(x, \cdot)$ belongs to $\Omega$. Since $B$ is orthogonalizing, $\delta_x(x) = 1$. Since the model is sharp  and state-complete, $\delta_x$ is the unique state with this property, and hence, pure. Since the model is state-complete, bi-symmetry guarantees that any pure state $\epsilon$ on $\E$ has the form $g\delta_x = n\B(g^{-1}x, \cdot)$ for some $g \in G$. Every extremal vector $v \in \E^{+}$ with $\B(v, u) = 1$ 
corresponds to a unique pure state $\epsilon_v$ via $\epsilon_v = \B(v, \cdot )$. Since $\B$ is non-degenerate,
it follows that $v = n g^{-1} x$ for some $g \in G$. But then $v \in \E_+$. It follows that $\E^{+} \subseteq \E_+$. 
$\Box$ 

Thus, if there exists an orthogonalizing SPIN {\em inner product} on $\E(A)$ ($A$ sharp, state-complete, and bi-symmetric) then $\E(A)$ is self-dual. In \cite{Wilce11b}, the existence of a 
{\em minimizing} SPIN inner product was simply postulated. Most of the remainder of this paper is devoted to finding reasonable sufficient conditions for the existence of such an inner product.

%{\em Question:} Can we replace transitivity of $G$ on pure states, by the double-sharpness condition?

%{\bf Lemma 3:} {\em Let $A = (X,\A,\Omega,G)$ be any fully symmetric model. Suppose $\langle, \rangle$ is an orthogonalizing SPIN inner product on $\E$. Let $\delta_x := n \langle x |$, and let $\Omega'$ be the closed convex hull of $\{ \delta_x | x \in X\}$ in $\Omega$. Let $A' := (X,\A,\Omega',G)$. This is a sharp, fully symmetric probabilistic model. Hence, $\E' := \E_{A'}$ self-dual.} 

%{\em Proof:} Note that $g \delta_x = \delta_{gx}$ for every $g \in G$, so $\Omega'$ is $G$-invariant. We need only show that $\Omega'$ is separating and sharp. The former is clear, as $X$ spans $\E$. For the latter, suppose $\delta_x (y) = 1$ for some $y \in X$. Then $1 = n \langle x, y \rangle = n r^2 \cos(\theta)$ where $\theta$ is the angle between $x$ and $y$. As $\langle, \rangle$ is orthogonalizing, $r^2 = 1/n$. Hence, $\cos(\theta) = 1$, whence, $\theta = 0$, and $y = x$. $\Box$

%So: any fully symmetric $G$-test space admitting an orthogonalizing representation, admits a self-dual (and also orthogonalizing) representation.

{\bf 2.3 Irreducible Systems} 

By the Corollary to Lemma 1, the ortho-complement $u^{\perp} = \{ x \in \E | B(x,u) = 0\}$ is independent of the SPIN form $B$.  I'll say that the model $A$ {\em irreducible} in case $u^{\perp}$ has no non-trivial $G(A)$-invariant subspace.\footnote{Admitting that this is again lousy terminology.} Things work especially nicely when $A$ is irreducible in this sense.

{\bf Lemma 3:} {\em Let $A$ be irreducible, and suppose $\B$ is any particular non-degenerate symmetric, invariant, normalized 
bilinear form on $\E$. Then all SPIN forms on $\E$ have the form 
\[\B_{\lambda}(a,b) := \lambda \B(a,b) + (1 - \lambda) \B(a,u)\B(u,b)\]
for some real parameter $\lambda$.} 

By the remark following Corollary 1, regardless of the choice of $\B$, we have $\B_o$ the uniform SPIN form, in conformity with our earlier usage.

{\em Proof:} Let $\B'$ be any SPIN bilinear form. Since $\B$ is non-degenerate, we have an 
operator $\beta : \E \rightarrow \E$, self-adjoint with respect to $\B$, such that $\B'(a,b) = \B( \beta a, b)$ for all $a, b \in \E$. Since $\B$ is $G$-invariant, $\beta$ is $G$-equivariant, i.e., $\beta (ga) = g\beta(a)$ for 
all $a \in \E$ and all $g \in G$. 

Let $u^{\perp}$ be the orthocomplement of $u$ with respect to $\langle, \rangle$. By Corollary 1, 
\[u^{\perp} = \{ a \in \E | \B'(a,u) = 0\} = \{ a \in \E | \B(a,u) = 0\},\] 
so $u^{\perp}$ is invariant under $\beta$. Let $\beta_o$ denote the restriction of 
$\beta$ to $u^{\perp}$, noting that this is still self-adjoint and $G$-equivariant. In particular, $\beta_o$ 
has a real eigenvalue $\lambda$, and the eigenspace $V_{\lambda}$ of $\beta_o$ is an invariant subspace of $u^{\perp}$. 
%(Indeed, if $v \in u^{\perp}$ is an eigenvalue for $lambda$, then $\beta(gv) = g\beta(v) = g \lambda v = \lambda gv = \lambda %gv$,) 
Since the latter is irreducible, and $V_{\lambda} \not = 0$ (by the non-degeneracy of $\B$), $V_{\lambda} = V$.\footnote{This is 
simply  the real form of Schur's Lemma.} Thus, for all $a_o, b_o \in u^{\perp}$, we have 
\[\B'(a_o,b_o) = \lambda \B(a, b).\]
If $a, b$ are now arbitrary vectors in $\E$, we can write $a = a_o + a_1$ and $b = b_o + b_1$, where 
$a_o, b_o \in u^{\perp}$ and $a_1 = \B(a, u)u$, $b_1 = \B(b,u) u$. By Corollary 1, we have 
$\B'(a_o,u) = \B(a_o, u) = 0$, so that 
$\B'(a_o,b_1) = \B(b, u) \B'(a_o,u) = 0$; likewise, $\B'(a_1,b_o) = 0$. Thus,  
\[\B'(a,b) = \B'(a_o + a_1, b_o + b_1) = \B(a_o,b_o) + \B(a_1,b_1) = \lambda \B'(a_o,b_o) + \B(a, u)B(b,u)\]
Since $a_o = a - B(a, u)u$ and $b_o = b - \B(b, u)u$, we can also write this as 
\[\B'(a,b) = \lambda (\B(a, b) - 2 \B(a, u)\B(b, u) + \B(a,u)\B(b,u))) + \B(a, u) \B(b, u).\]
Simplifying, this gives us $\B'(a,b) = \lambda \B(a, b) + (1 - \lambda)\B(a,u)\B(b,u)$, 
as promised. $\Box$ 

%Notice that (minima achieved at same place? Later??)

%{\bf Corollary 4:} {\em If $A$ is irreducible, then if one non-degenerate SPIN form on $\E(A)$ is minimizing, all are 
%minimizing.} 

%{\bf Corollary 2:} {\em Let $u^{\perp}$ be irreducible, as in Lemma 4. Then, for any SPIN bilinar form $B$ on $\E$, 
%$B(a,u) = \langle a, u \rangle$ for all $a \in \E$. In particular, $u^{\perp} = \{ a \in \E | B(a,u) = 0\}$.}

%{\em Proof:} Suppose that $B = B_{\lambda}$. Then 
%\[ \ \ B(a,u) = \lambda \langle a, u \rangle + (1-\lambda)\langle a, u \rangle = \langle a, u \rangle. \ \Box\]

Suppose now that $c_{\lambda} = \B_{\lambda}(x,y)$ for orthogonal $x,y$, and let $c = \B(x,y)$ where $\B$ is a chosen SPIN {\em inner product} (say, the standard one 
arising from group averaging). Then 
\[c_{\lambda} = \lambda (c - 1/n^2) + 1/n^2 = \lambda \frac{n^2 c - 1}{n^2} + \frac{1}{n^2}.\]
which is $0$ iff 
\[\lambda = \frac{1}{n^2} \left ( \frac{n^2}{1 - n^2 c} \right ) = \frac{1}{1 - n^2c}.\]
(Notice that $1 - n^2c = 0$ only for $c = 1/n^2$, which is to say, only if $\B = \B_o$, which we have ruled out by taking $B$ to be an inner product, so this value of $\lambda$ is legitimate.) 
It follows that there is {\em at most one} orthogonalizing SPIN form on $\E$, this corresponding to a non-negative value of $\lambda$. In order to guarantee that such a form exists, 
we need to know something more about the positivity of the forms $\B_{\lambda}$.  As above, let $\B$ be any chosen SPIN inner product on $\E$; as in Lemma 1, let $m$ denote the minimum value of $B(x,y)$ as $x,y$ range over $X$, and recall that, by part (d) of Lemma 1, this never exceeds $1/n^2$. 
 
{\bf Lemma 4:} {\em Let $A$ be irreducible, with $\dim \E(A) > 1$. Let $\B$ denote any particular SPIN inner product on $\E$ (say, the standard one arising from group averaging) on $\E$, and 
let $m$ and $M$ denote, respectively, the minimum and maximum values of $\B$ on $X \times X$. Then, for any $\lambda \in {\Bbb R}$, 
\begin{itemize} 
\item[(a)] The minimum value $m_{\lambda}$ of $\B_{\lambda}$ is given by 
\begin{equation} m_{\lambda} = \left \{ \begin{array}{lcl} 
\lambda m + (1-\lambda)/n^2 \ = \ \lambda(m - 1/n^2) + 1/n^2 & \text{if} & \lambda \geq 0 \\
 &  & \\
 \lambda M + (1 - \lambda)/n^2 \ = \ \lambda(M - 1/n^2) +  1/n^2 & \text{if} & \lambda < 0 \end{array} \right .\end{equation}
\item[(b)] $\B_{\lambda}$ is  positive iff 
\[\frac{1}{1 - Mn^2} \leq \lambda \leq \frac{1}{1 - mn^2}\]
where $m$ and $M$ are, respectively, the minimum and maximum values of $\B$ on $X \times X$.
\item[(c)] $\B_{\lambda}$ is positive-semidefinite iff $\lambda \geq 0$, and an inner product iff $\lambda > 0$. 
\end{itemize}} 

{\em Proof:} (a) If $\lambda$ is non-negative, the minimum of $\lambda \B(x,y) + (1 - \lambda)/n^2$ occurs where $\B$ is minimized; if $\lambda < 0$, the minimum occurs where $\B$ is maximized. (b) From part (a), we see that for $\lambda \geq 0$, $m_{\lambda}$ is non-negative iff $\lambda (m - 1/n^2) \geq -1/n^2$, 
or, equivalently, (recalling that $m < 1/n^2$, so that $m - 1/n^2$ is negative) 
\[\lambda \leq -\frac{1}{n^2(m - 1/n^2)} = \frac{1}{1 - mn^2}. \]
Similarly, if $\lambda < 0$, $m_{\lambda} \geq 0$ iff $\lambda \geq 1/(1 - Mn^2)$. 
(c) Now suppose that $\B_{\lambda}(a,a) \leq 0$. Then  $\lambda \|a\|^2 + (1 - \lambda)\B(a, u)^2 < 0$. Since $u$ is normalized, and $\B$ is an inner product, we have $\B(a,u)^2 \leq \|a\|^2$. 
%\footnote{Since $\B$ is an inner product, we have $|\B(a, u)| \ \|a\|^2 \cos^2 \theta$ where $\theta$ is the angle between $a$and $u$}, 
whence, as $\B(a, u)^2 \geq 0$, we must have $\lambda \leq 0$. Thus, if $\lambda > 0$, $\B_{\lambda}$ is positive-definite, that is, an inner product. Conversely, suppose $\lambda \leq 0$. Let
$r^2 = \B(x,x) \leq 1/n^2$ and $r_{\lambda}^{2} = \B_{\lambda}(x,x)$, and recall that, since 
$\B$ is an inner product, Lemma 1 (d) gives us 
$r^2 > 1/n^2$ (the inequality strict, as $\E$ is not one-dimensional.) Now we have 
$r_{\lambda}^{2} = \B_{\lambda}(x,x) = \lambda (r^2 - 1/n^2) + 1/n^2 < 1/n^2$. But now, again by Lemma 1 (d), $\B_{\lambda}$ is not an inner product. $\Box$ 

It now follows that, at the critical value $\lambda = 1/(1 - cn^2)$ for which the form $\B_{\lambda}$ is orthogonalizing, $\B_{\lambda}$ is positive --- a SPIN form --- iff $\lambda \leq 1 / (1 - mn^2)$, i.e, 
\[(1 - mn^2) \geq (1 - cn^2) \ \Leftrightarrow \ -mn^2 \leq -cn^2 \ \Leftrightarrow \ c \leq m\]
which occurs iff $c = m$, i.e., iff $\B$ is minimizing. Thus, we have

{\bf Corollary 2:} {\em Let $A$ be irreducible. Then $\E$ supports an orthogonalizing SPIN form iff it supports a minimizing SPIN form. In this case,  the {\em unique} orthogonalizing SPIN form is the form   $\B_{\overline{\lambda}}$, where $\overline{\lambda} = \frac{1}{1 - mn^2}$, the maximum value of $\lambda$ for which $\B_{\lambda}$ is positive. This is an inner product.} 

%BETTER: renormalize in terms of $1$. 

%With $B$ fixed as above, let $\overline{\lambda} = 1/(1 - mn^2)$ and $\underline{\lambda} = 1/(Mn^2 - 1)$ denote the maximum %and minimum values of the parameter $\lambda$ for which $B_{\lambda}$ is positive, as established in 
%part (b) of Lemma 5. 

%{\bf Corollary:} {\em With notation as above, $m_{\overline{\lambda}} = m_{\underline{\lambda}} = 0$.} 

%{\em Proof:}  Solve the equations. 

%It follows that $B_{\overline{\lambda}}$ and $B_{\underline{\lambda}}$ are the {\em only candidates} for an orthogonalizing %SPIN form on $\E$. However, 

%{\bf Corollary:} {\em  In particular, then, {\em if a SPIN form is orthogonalizing} on an irreducible system, 
%{\em it must be an inner product}. 

We are free to replace the given SPIN inner product $\B$ in Lemma 2 with any other SPIN inner product. Choosing the SPIN inner product $\B_{\overline{\lambda}}$ for the maximal value 
$\overline{\lambda}$, we obtain a range of values $0 \leq \lambda \leq 1$. \underline{\em Henceforth, I assume this parametrization}, so that $\overline{\lambda} = 1$.  
To emphasize that $\B_{1}$ is an inner product, I'll sometimes write it as $\langle , \rangle_1$. 

It follows that, where $A$ is irreducible, the inner product $\B_1 = \langle \ , \ \rangle_1$ is the {\em only candidate} for an orthogonalizing SPIN inner product. To put it another way: if $A$ is irreducible, then there exists at most one  orthogonalizing SPIN form on $A$, {\em and this is an inner product}.

\section{Composites and Conjugates} 

Evidently, what is now wanted is a phyically (or operationally, or probabilistically) natural condition guaranteeing the existence of an orthogonalizing (equivalently, minimizing) SPIN form. In this section I will offer three (not entirely independent) such conditions. All turn on the notion of a composite system. Very briefly: there is a correspondence between {\em non-signaling} bipartite states and bilinear forms, so that equivariant bipartite states give rise to SPIN forms. The game is to seek conditions on such a state that (i) have a clear physical (or operational, or probabilistic) meaning, and (ii) 
guarantee that the corresponding SPIN state is orthogonalizing. I'll start with a quick review of how composite systems are handled in the current framework. More detail can be found in 
\cite{BBLW, Wilce10}.\footnote{Another, unrelated, motivation is sketched in \cite{Wilce11b}. By choosing a fixed pure state $\epsilon_o$, we can represent elements of $\E$ as continuous random variables on $G$, via $x \in X \mapsto \hat{x} \in {\Bbb  R}^{G}$, where $\hat{x}(g) = \alpha(g x)$. That the 
canonical inner product obtained by group averaging be minimizing --- which, in view of Corollary 2, is equivalent to the existence of an orthogonalizing form, at least for irreducible models, is equivalent to the condition that the covariance $\cov(\hat{x},\hat{y})$ of two of these random variables be minimized precisely when the corresponding outcomes are distinguishable.} \\  
%[Compare w/ earlier...]

%MORPHISMS HERE? COMP as morphism?

%By a {\em monoidal probabilistic theory}, I mean something a bit more specific than a category of models equipped with a %symmetric monoidal structure. Rather, I mean such a set-up, but with the additional requirement that, 
%for two models $A, B \in \C$, the monoidal product $A \otimes B$ be a {\em composite}, in the following 
%sense \cite{W11}:

%\newpage
{\bf 3.1 Composite systems and non-signaling states}

{\bf Definition 11 (Composites):} A {\em composite} of two models $A$ and $B$ is a model 
$AB$, plus an injection  $X(A) \times X(B) \rightarrow X(AB)$, which I'll write as $(x,y) \mapsto xy$, 
such that 
\begin{itemize}
\item[(i)] for all $E \in \A(A)$ and $F \in \A(B)$, $EF := \{ xy | x \in E, y \in F \} \in \A(AB)$, 
\item[(ii)] for all $\alpha \in \Omega(A)$, $\beta \in \Omega(B)$, there exists some $\gamma \in \Omega(AB)$ with 
$\gamma(xy) = \alpha(x)\beta(y)$; and 
\item[(iii)] for all $g \in G(A), h \in G(B)$, there exists some $k \in G(AB)$ with 
$k(xy) = (gx)(hy)$ for every $x \in X(A), y \in X(B)$. \footnote{The notation $AB$ is not to be understood as referring (yet) to any particular {\em operation} of composition. That is, $AB$ does not refer --- as yet, anyway --- to any {\em particular} composite.}
 
\end{itemize} 

Condition (i) of the definition allows us to identify $X(A) \times X(B)$ with the set $X(A)X(B) = \{ xy | x \in X, y \in Y\}$ of {\em product outcomes} in $X(AB)$. Let us write $\A(A) \times \A(B)$ for the set of {\em product tests}, i.e., tests of the form $EF$ provided for by condition (i). Evidently, every state in $\Gamma$ restricts to a state on $\A \times $; by (ii), the set of such restrictions 
contains all {\em product states} $\alpha \otimes \beta$, defined by $(\alpha \otimes \beta)(xy) = \alpha(x) \beta(y)$. Also, by (iii), the stabilizer in $G(AB)$ of the set $X(A)X(B)$ extends the action on the latter of $G(A) \times G(B)$. 

Another consequence of condition (i) is that 
\[x_1 \perp x_2 \ \text{in} \ X(A) \ \ \Rightarrow \ \ x_1 y \perp x_2 y \ \text{in} \ X(AB)\]
for every choice of $y \in X(B)$; likewise, if $y_1 \perp y_2$ in $X(B)$, then $x y_1 \perp x y_2$ for every $x \in X(A)$. 
This observation will be exploited below. 

%[Examples...]

{\em Remark:} The category of {\em all} probabilistic models and morphisms has a natural product structure. Given 
models $A$ and $B$, let $A \times B$ be the model with outcome set $X(A) \times X(B)$, test space 
$\A(A) \times \A(B) = \{ E \times F | E \in \A(A), F \in \A(B)\}$, state space the convex hull of the product states, and symmetry group $G(A) \times G(B)$, acting as usual. (This is {\em not} a cartesian structure, since there are in general no morphisms $A \times B \rightarrow A$ to serve as projections.) If we strengthen 
condition (ii) in Definition 11 to require that there exist a group homomorphism  $\psi : G(A) \times G(B) \rightarrow G(AB)$ with $\psi(g,h)(xy) = (gx)(gy)$, then $g, h \mapsto k$ is a homomorphism, and  $x,y \mapsto xy$ defines a morphism $A \times B \rightarrow AB$. 

%{\em Remark:} The definition of composite is not very restrictive. For instance, if $A$ and $B$ are two models, then any model %of the form 
%$(X(A) \times X(B), \A(A) \times \A(B), \Omega(A) \mintensor \Omega(B), G(A) \times G(B))$ is a composite (where $\A(A) \times %\A(B)$ is the set of product tests $E \times F$, $E \in \A(A)$, $F \in \A(B)$, and $\Omega(A) \mintensor \Omega(B)$ is the %closed convex hull of the product states. If $A(\H)$ and $A(\K)$ are quantum models, one takes 
%$A(\H) A(\K) = A(\H \otimes \K)$, in which case the composite state space lies properly between the maximal and 
%minimal products.  
%$
%%However, we could replace the latter by $\Omega(\A(A) \times \A(B))$, the set of all 
%probability weights on $\A(A) \times \A(B)$. 

A state $\omega$ on a composite systenm $AB$ is {\em non-signaling} \cite{KFR} iff it has well-defined {\em marginal} (or reduced) states 
\[\sum_{x \in E} \omega(xy) =: \omega_2(y) \ \ \text{and} \ \ \sum_{y \in F} \omega(xy) =: \omega_1(x),\]
independent of the choice of tests $E \in \A(A)$, $F \in \A(B)$. In this case, for every $y \in X(B)$ and $x \in X(A)$, we define the {\em conditional states} $\omega_{1|y}$ and $\omega_{2|x}$ on $(X,\A)$ and $(Y,)$, respectively, by 
\[ \omega_{1|y}(x) := \frac{\omega(xy)}{\omega_2 (y)} \ \text{and} \ \omega_{2|x}(y) := \frac{\omega(xy)}{\omega_1(x)}.\]
It is straightforward to establish the following bipartite {\em laws of total probability} for a non-signaling state $\omega$: 
\begin{equation} \omega_1  \ = \ \sum_{y \in F} \omega_2(y) \omega_{1|y} \ \ \text{and} \ \ \omega_2 =  \sum_{x \in E} \omega_1(x) \omega_{2|x}\end{equation} 
for any choices of tests $F \in \A(B)$ and $E \in \A(A)$.

{\bf Definition 12:} A composite $AB$ of models $A$ and $B$ is non-signaling iff all of 
its states are non-signaling, {\em and} all conditional states belong to the designated state spaces of $A$ and $B$ --- that is, 
$\omega_{2|x} \in \Omega(B) \ \ \text{and} \ \ \omega_{1|y} \in \Omega(A)$ for all $x \in X(A)$ and $y \in X(B)$.

In particular, then, if $AB$ is a non-signaling composite in the sense just 
defined, then $\omega_1 \in \Omega(A)$ and $\omega_2 \in \Omega(B)$ for every state $\omega \in \Omega(AB)$. 
%[LT -- comments here?]

It is not hard to show (see \cite{Wilce92}) that if $\omega$ is non-signaling, then it gives rise to a unique bilinear form $\B_{\omega}$ on $\E(A) \times \E(B)$ with 
$\B_{\omega}(x,y) = \omega(x,y)$ for all outcomes $x \in X(A), y \in X(B)$. \footnote{Conversely, if $\omega$ is given by a biliner form $\B_{\omega}$, it must be non-signaling, since we then have 
\[\sum_{x \in E} \omega(x,y) = \sum_{x \in E} \B_{\omega}(x,y) = \B_{\omega}(u,y)\]
for all $E \in \A$, and similarly in the second argument.}
Thus, every  non-signaling state $\omega$ on $AB$ is associated with a positive linear mapping $\E(A) \rightarrow \E(B)^{\ast}$, given by 
$\widehat{\omega}(a)(b) = \B_{\omega}(a)(b)$ for all $a \in \E(A), b \in \E(B)$. Since the conditional states 
$\omega_{2|x}$ and $\omega_{1|y}$ lie in $\V(A)$ and $\V(B)$, respectively, the range of this mapping is 
contained in $\V(B)$, so we can --- and I shall --- regard $\widehat{\omega}$ as a positive linear mapping 
\[\hat{\omega} : \E(A) \rightarrow \V(B).\] If $\langle , \rangle$ is a self-dualizing inner product 
on $\E(B)$, we can re-interpret $\widehat{\omega}$ as positive linear mapping 
$\widehat{\omega} : \E(A) \rightarrow \E(B)$, given by the condition 
\[\langle \widehat{\omega}(a),b \rangle = B_{\omega}(a,b).\]
Notice that $\widehat{\omega}^{\ast}(b)(a) = \widehat{\omega}(a)(b) = \omega(a,b)$. Notice, too, that $\widehat{\omega}(u)(y) = \sum_{x \in E} \omega(x,y)$, i.e., 
$\widehat{\omega}(u)$ is the marginal of $\omega$ in $\Omega(X(B),\A(B))$, whence, 
$\widehat{\omega}(x)/u(\widehat{\omega}(x))$ is just the conditional state $\omega_{2|x}$. Accordingly, $\widehat{\omega}$ is called the 
{\em conditioning map} associated with $\widehat{\omega}$  Where this map is an order-isomorphism $\E(A) \rightarrow \V(B)$, we say that $\omega$ is an {\em isomorphism state} \cite{BGW}. 

%Where $\E(B)$ is self-dual, this lies is $\E(B)_+$, i.e., is an element of $\Omega(B)$.  

{\em Jargon:} Let $\omega$ be a non-signaling state on $AA$, and let $\B_{\omega}$ be the corresponding bilinear form. If $\B_{\omega}$ is a SPIN form, I'll call $\omega$ a SPIN state. 
%if $\B_{\omega}$ is the uniform 
%SPIN form, I'll refer to $\omega$ as the uniform state. 

%{\bf [Note: no restriction on marginals/conditionals as yet...!!]}
 
A trivial but important example of a non-signaling state is the {\em uniform} (or {\em maximally mixed}) state on $AA$: $\rho(z) = 1/n^2$ for every $z \in X(AA)$. The associated SPIN form, with parameters $c = r^2 = 1/n^2$, is exactly the degenerate, or uniform, SPIN form $\B_o$.  %(If $A$ is irreducible, then by Corollary ...., this is the ony state with these parameters.)

%{\bf BIG QUESTION HERE: Is $B(x,y) \equiv 1/n^2$ the UNIQUE SPIN form with $c = r^2 = 1/n^2$?? [Among pos-sd, yes.] [Also 
%for irreps!]}

For later reference:

{\bf Definition 13 (Local Tomography):} A composite $AB$ is {\em locally tomographic} iff bipartite states in $\Omega(AB)$ are uniquely determined by their values on product outcomes --- that is, iff
for all $\omega_1, \omega_2 \in \Omega(AB)$, 
\[\omega_1(x,y) = \omega_2(x,y) \forall x \in X(A), y \in X(B) \ \Rightarrow \ \omega_1 = \omega_2.\]

{\em Remark:} If $AB$ is non-signaling, then (in our current, finite-dimensional setting), local tomography sets up a linear (NB: not ordered-linear) isomorphism $\E(AB) \simeq \E(A) \otimes \E(B)$. The 
cone on $\E(A) \otimes \E(B)$ obtained by carrying forward the cone $\E_+(AB)$ sits between the minimal (or projective) cone generated by the product states, and the maximal (or injective) cone consisting of all positive bilinear forms on $\E^{\ast}(A) \otimes \E^{\ast}(B)$ \cite{BBLW, Wilce92}.

Both non-signaling and local tomography conditions are routinely assumed (sometimes explicitly, sometimes tacitly) in recent discussions of composite systems in generalized probabilistic theories (\cite{Hardy, BBLW07, Rau, CDP}, etc.). The non-signaling condition will be important in what follows, but the extremely powerful local tomogrcaphy assumption plays no role here at all (but see further comments in the Conclusion). \\

%\newpage
%PUT DISCUSSION OF CONJUGATES HERE!?
{\bf 3.2 Conjugate Systems}

In view of the fact that equivariant non-signaling states yield SPIN forms, it is temping simply to {\em postulate} the existence of a state $\omega$ on a composite $AA$ with the property that $\omega(xy) = 0$ for all $x \perp y$ in $X(A)$. Such a state would {\em perfectly correlate} every test $E \in \A$ with itself, in that, where Alice and Bob perform the same test at their locations, they are guaranteed the same outcome. 

Unfortunately, in ordinary quantum theory, {\em there is no such state}: the candidate is the normalized trace, i.e., 
$B(x, y) = \Tr(P_x P_y) = |\langle x, y \rangle|^2$, 
which corresponds to no bipartite density operator. Fortunately, though, the strategy {\em does} work with a small modification. Consider a complex Hilbert space $\H$ and its 
conjugate space $\overline{\H}$, \
%and consider $\H \otimes \bar{\H}$, 
%with its standard inner product, $\langle x \otimes y, u \otimes v \rangle = \langle x, u \rangle \langle v, y \rangle$. Let 
and let $\Psi \in \H \otimes \bar{\H}$ be the (twisted?) Bell state 
\[\Psi = \sum_{x \in E} x \otimes \bar{x}\]
where $E$ is any orthonormal basis. This is independent of the chosen basis, and perfectly correlates every observable with its conjugate analogue --- indeed, $\langle \Psi, x \otimes \bar{y} \rangle = \langle x, y \rangle$, so that 
$|\langle \Psi, x \otimes y \rangle |^2 = |\langle x, y \rangle |^2.$

This suggests the following idea. %Recalling the definition of a morphism of models from Section 1, we see that an 
Recall from Section 1 that an {\em isomorphism} from a model $A$ to a model $B$  consists of a bijection $\phi : X(A) \rightarrow X(B)$ taking $\A(A)$ bijectively onto $\A(B)$, and such that $\beta \mapsto \beta \circ \phi$ is an affine isomorphism from $\Omega(B)$ to $\Omega(A)$, {\em plus} an action of $G(A)$ on $B$ by elements of $G(B)$, such that  isomorphism $\psi : G(A) \rightarrow G(B)$ such that $\phi(gx) = g\phi(x)$ for all $x \in X(A)$, $g \in G(A)$. %[This should go earlier, AND we'll need to remark that morphisms lift to linear maps etc.]

{\bf Definition 14 (Conjugate Models):} A {\em conjugate} for a model $A$ is a structure $(\overline{A}, \gamma_A, \eta_A)$, where 
$\overline{A}$ is a model, $\gamma_A : A \rightarrow \overline{A}$ is an isomorphism, and 
$\eta_A$ is a bipartite state (on some non-signaling composite) $A \overline{A}$ such that 
\[\eta_{A}(x,\gamma_{A}(x)) = 1/n\]
for every $x \in X(A)$. I'll call $\gamma_A$ the {\em conjugation map} and $\eta_{A}$, the {\em correlator} for the given conjugate.

%[NOTE: we need to be careful about the difference between $\eta : \E(AB) \rightarrow {\Bbb  R}$ and 
%the pull-back $\eta \circ \otimes : \E(A) \times \E(A) \rightarrow {\Bbb  R}$. That is, point out that 
%any state on $AB$ induces a bilinear form on $\E(A) \times \E(B)$. 

%[NOTE: how far can we get without the uniformity assumption? What work does that do, absent uniqueness?] 

%Observe that $\langle a, b \rangle := \eta(a, \gamma_{A}(b))$ is then an orthogonalizing SPIN form on $A$. 
%So, if $A$ is irreducible, it's an inner product, by Lemma ...  . [Here, might need to symmetrize?]

{\bf Example 7: Quantum Cases} If $A = A(\H)$ is a quantum model associated with a complex Hilbert space $\H$, 
let $\overline{A} = A(\bar{\H})$; let $\gamma_{A} : X(\H) \rightarrow X(\bar{\H})$ be the mapping $x \mapsto \bar{x}$ 
(strictly speaking, the identity map!), and let $\eta_{A}(x,\gamma_A(y)) = |\langle \Psi, x \otimes y \rangle|^2 
= \Tr(P_{\Psi}P_{x \otimes y})$. As discussed above, this last is a correlator --- obviously, symmetric and 
invariant.

{\bf Lemma 5:} {\em If $A$ has a conjugate, then it has a conjugate for which the correlator $\eta_A$ is symmetric, 
in the sense that $\eta(x,\gamma_A(y)) = \eta(y, \gamma_A(x))$, and invariant, in the sense that $\eta_{A}(gx, \overline{g}\overline{y}) = \eta(x,\overline{y})$.} 

{\em Proof:} Let $\eta^{T}(x,\gamma_{A}(y)) := \eta(y, \gamma_{A}(x))$. Observe that this is again a correlator. Averaging 
the two gives us a symmetric correlator. Now suppose $\eta$ is symmetric, and consider $\eta^{g}(x,y) = \eta(gx, gy)$. This 
again is a symmetric correlator, so averaging over the group yields an invariant symmetric correlator. $\Box$

Convention: Henceforth, assume that correlators are symmetric and invariant. It follows that 
\[\B(a,b) := \eta(a, \gamma_{A}(b))\] 
is an orthogonalizing SPIN form  --- and hence, if $A$ is irreducible, an orthogonalizing SPIN inner product --- on $\E(A)$. 
%Notice, too,  that if $A$ is irreducible, $\widehat{\eta} : \E(A) \rightarrow \V(\bar{A})$ is an equivariant map --- hence, by irreducibility of $A$ (and Schur's Lemma), a linear isomorphism.  %%CHECK!

%%\newpage
{\bf Theorem 1:} {\em Let $A$ be  bi-symmetric, and have a conjugate $(\bar{A}, \gamma_A, \eta_A)$. Then the following are equivalent:
\begin{itemize}
\item[(a)] $A$ is state-complete and $\eta$ is an isomorphism-state.
\item[(b)] $A$  is self-dual with respect to the form $\B(a,b) := \eta_{A}(a,\gamma(b))$. 
\end{itemize} 
If $A$ is irreducible, then $B$ is an inner product, and (a) and (b) are equivalent to 
\begin{itemize} 
\item[(c)] $A$ is state-complete and sharp. 
\end{itemize} } 

{\em Proof:} (a) $\Rightarrow$ (b): If $\eta$ is an order-isomorphism, $\widehat{\eta}^{\ast}$ takes $\E_+$'s extremal rays to those of $\V_{+}$. Since $A$ is state-complete, the latter is $\E^{\ast}_{+}$. In particular, $\hat{\eta^{\ast}}(\bar{x}) = \eta_2(\bar{x}) \eta_{1|\bar{x}}$ is pure, and every pure state {\em on} $\E(A)$ looks like this. We therefore have transitivity of $G$ on pure states of $\bar{A}$, and also that $\widehat{\eta}^{\ast}(\bar{x})(y) = \B(y,x) = \B(x,y)$, so that 
$\widehat{\eta}_{1|\bar{x}} = n\B(x,\cdot)$ corresponds to a point in $\E_+$. Thus, $\E^{+} \subseteq \E_+$.

(b) $\Rightarrow$ (a) Conversely, suppose $A$ is self-dual with respect to $\B$. Let $\tau = \widehat{\eta}^{\ast} \circ \gamma : \E \rightarrow \V = \E^{\ast}$ (the latter identity, one of linear 
spaces, not yet of ordered linear spaces). We have 
$\tau(\E_+ ) \subseteq \V_+ \subseteq \E^{\ast}_+$, so this is a positive mapping. Since 
$\E$ is self-dual with respect to $\B$, we have $\ker(\tau) \leq \E^{+} = \E_+$; since the latter 
cone contains no subspaces other than $0$, $\tau$ is injective, and thus, in the present finite-dimensional 
setting, a linear isomorphism. 
%As remarked above, $\widehat{\eta}$ is a linear isomorphism; hence, so is $\tau$. 
%Thus %Why "thus"?
The definition of $\tau$ gives us $\B(a, b) \ = \ \tau(b)(a)$ 
for all $a, b \in \E(A)$. Thus, for $\beta = \tau(b)$, $b \in \E$, we have 
\[\beta \in \E^{\ast}_+ \ \Leftrightarrow \ \tau(b)(a) \geq 0 \ \forall a \in \E_+ \ \Rightarrow b \in \E_+ \]
(the last, by self-duality), whence, $\tau^{-1} (\E^{\ast}_+) \subseteq \E_+$. In other words, $\tau$ is an order-isomorphism. Since $\gamma$ is also such, it follows that 
$\widehat{\eta}^{\ast}$ is an order-isomorphism, i.e., $\eta_A$ is an isomorphism state. Moreover, we have 
\[\E^{\ast}_+  = \tau(\E^{+}) = \tau(E_+) \subseteq \V_+,\]
whence, $\E^{\ast}_+  = \V_+$, i.e., $A$ is state-complete. %[WHAT ABOUT SHARP??] 

Suppose now that $A$ is irreducible. Corollary 2 then tells us that the orthogonalizing SPIN form $\B$ is an inner product. It follows from Lemma 2 that (c) $\Rightarrow$ (a). 

(b) $\Rightarrow$ (c): We saw above that (b) implies state-completeness. Since $A$ is irreducible, the orthogonalizing SPIN form $\B$ is an inner product (indeed, $\B = \B_1$, in the notation of Section 2.3). We have $\delta_x := n \langle x | \in \V^{+}$, so that $nx \in \E^{+} = \E_+$, with $\langle nx, x \rangle = 1$. Since $x$ is extremal in $\E_+$, $\delta_x$ is extremal in $\V_+ = \E^{\ast}_+$. By state-transitivity, every pure state has the form $\delta_y = n \langle y |$ for some $y \in X$. In particular, then, for every vector $v \in \E^{+}$ with $\langle v, x \rangle = \langle v, u \rangle = 1$, we have $\|v\| = \|nx\|$. It follows that 
if $\langle v, x \rangle = \langle x, x \rangle = 1$, $v = x$. Thus, $A$ is sharp. $\Box$ 

%(c) $\Rightarrow$ (a): Since $B(a,b) := \eta(a,\gamma(b))$ is an orthogonalizing SPIN form, this follows from Lemma 2 and Corollary 2. $\Box$  
 %[Converse? What about SC+sharp (or crisp)?]

%{\em Remark:} We seem to need irreducibility only for the equivalence of (b) and (c). So perhaps re-state?

%%%\newpage

%{\em Remark:} If the equivalent conditions of Theorem 1 are satisfied, then $A$ is state-complete (since $\E_+ \subseteq \vec{\V}_+ \subseteq \E^{+} = \E_+$, whence, $\V_+ = \E^{\ast}_+$) and 
%also sharp: $n\langle x |$ is the unique vector of unit length with $\langle v, x \rangle = 1$, and any pure state, being the image of this one under a symmetry, also has norm one. [Put these remarks 
%earlier, under a discussion of what self-duality must imply, for a bi-symmetric model. Also: introduce $\vec{\alpha}$ notation earlier.
Where a correlator $\eta$ is an isomorphism state, I'll call it an {\em iso-correlator}. Using this jargon, we have 

{\bf Corollary 3:} {\em Let $A$ be state-complete, bi-symmetric, irreducible, and have a conjugate with an iso-correlator. Then $A$ is self-dual.} 
%(One might reasonably tighten the definition of a conjugate to require that the correlator be an isomorphism state.)

In \cite{Wilce10}, I called a bipartite state $\omega \in AB$ on two rank-$n$ test spaces {\em correlating} 
iff, for some pair of tests $E \in \A(A)$ and $F \in \A(B)$, there exists a bijection $f : E \rightarrow F$ such that $\omega(x,f(y)) = 0$ for $x \not = y$. Evidently, the correlator of a conjugation is correlating in this sense (choose $E \in \A(A)$ and $F = \gamma_{A}(E) \in \A(\bar{A})$, and let $f(x) = \gamma_{A}(x)$ for $x \in E$). The {\em correlation condition} of \cite{Wilce10, Wilce11b} requires that every state on $A$ arise as the marginal of some correlating bipartite state on a composite of two copies of $A$.  

A stronger condition than the existence of a conjugate system, which will turn out to be useful, is the following 

{\bf Definition 15 (Strong Conjugates):} A  {\em strong conjugate} for a model $A$ consists of 
a system $\bar{A}$, an isomorphism $\gamma_{A} : A \simeq \bar{A}$, and a composite $A\bar{A}$, such that 
for every state $\alpha \in \Omega(A)$, there exists a non-signaling state $\omega^{\alpha} \in \Omega(A\bar{A})$ satisfying   
\begin{itemize} 
\item[(a)] $\omega^{\alpha}_1 = \alpha$ (that is, $\omega^{\alpha}$ is a dilation of $\alpha$)  %[BTW, introduce "dilation" earlier...]
\item[(b)] $\omega^{\alpha}(gx, g\bar{y}) = \omega(x, \bar{y})$ for all $g$ fixing $\alpha$, and 
\item[(c)] $\omega^{\alpha}$ is {\em correlating along $\gamma_A$}, in the sense that there exists at least one test $E \in \A$ with 
$\omega(x,\bar{x}) = \alpha(x)$ for all $x \in E$ (where, as above, $\bar{x} = \gamma_{A}(x)$).
\end{itemize} 

%[NOTE: can make this symmetric, 
%if we wish...] 

Notice that a strong conjugate is (in effect) a conjugate, since we can take $\eta$ to be $\omega^{\rho}$, where $\rho$ is the uniform state on $A$. 
%[Questions: (1) Do we need/want the correlator to be part of the definition of a conjugate? (2) Do we 
%want to make the coposite part of the definition (in either case)?]

{\bf Example 8: The quantum case}. That the conjugate, $A(\bar{\H})$, of a quantum model $A(\H)$, is in fact a strong conjugate 
is essentially just the Schmidt decomposition. Let $\H$ be a Hilbert space, and, as above, let 
$\bar{\H}$ denote the conjugate Hilbert space. For $x, y \in \H$, let $x \odot y$ denote 
the  operator on $\H$ given by $(x \odot y) z = \langle z, y \rangle x$. In particular, if $x$ is a unit vector, 
then $x \odot x = P_x$, the orthogonal projection operator associated with $x$. 
The mapping $x, y \mapsto x \odot y$ is sesquilinear, that is, linear in its first, and conjugate linear in its second, argument. Hence, there is a natural linear isomorphism $\H \otimes \bar{\H} \simeq {\cal B}(\H)$ taking $x \otimes \bar{y}$ to $x \odot y$. Suppose now that 
$W$ is a density operator on $\H$, diagonalized by an orthonormal basis $E \in \A(\H)$. Then $W$ has spectral 
resolution 
\[W = \sum_{x \in E} \lambda_x P_x = \sum_{x \in E} \lambda_{x} x \odot x.\]
The corresponding vector in $\H \otimes \bar{\H}$ is then 
\[\Psi_{W} := \sum_{x \in E} \lambda_{x} x \odot \bar{x}.\]
If $u, v \in E$ with $u \perp v$, then for every $x \in E$, we have either $\langle x, u \rangle = 0$ or 
$\langle u, y \rangle = 0$, whence,  
\[\langle \Psi_{W}, u \otimes \bar{v} \rangle \ = \ \sum_{x \in E} \lambda_x \langle x, u \rangle \langle \bar{x}, v \rangle = 0.\]
Moreover, on the diagaonl, we have 
\[\langle \Psi_{W}, u \otimes \bar{u} \rangle = \sum_{x \in E} \lambda_{x} |\langle u, x \rangle|^2 = \langle W u, u \rangle.\]
Evidently, the pure state corresponding to $\Psi$ sets up a perfect correlation between $E$ and its corresponding 
test $\bar{E}$, along the canonical isomorphism $x \mapsto \bar{x}$. Of equal note, if $g$ is a unitary leaving 
$W$ fixed, i.e, with $gWg^{-1} = W$, then the bipartite state (corresponding to) $\Psi$ is also invariant 
under the diagonal action of $G = U(\H)$: %[explain and demonstrate...] 
\begin{eqnarray*}
\omega(gu, \bar{gv}) & = & \sum_{x \in E} \lambda_x \langle x \otimes \bar{x}, gu \otimes \overline{gv} \rangle\\ 
& = & \sum_{x \in E} \lambda_x \langle x, gu \rangle \langle \bar{x}, \bar{g}\bar{v} \rangle \\
& = & \sum_{x \in E} \lambda_x \langle g^{-1} x, u \rangle \langle \bar{g}^{-1} \bar{x}, \bar{v} \rangle \\
& = & \left \langle \left (\sum_{x \in E} \lambda_x g^{-1} x \otimes \bar{g}^{-1} \bar{x} \right), u \otimes \bar{v} \right \rangle
\end{eqnarray*} 
(where, for an operator $a$ on $\H$, $\bar{a}$ denotes the linear operator on 
$\bar{\H}$ given by $\bar{a}(\bar{v}) = \bar{av}$ for all $v \in \H$.)

{\bf 3.3 Factorizable States} 

Another way to motivate the existence of an orthogonalizing SPIN form on an irreducible system $A$ is to suppose 
there exists an irreducible system $B$ (perhaps another copy of $A$) and a non-trivial SPIN form $\B$ on a composite $AB$ that {\em factors}, in the following sense:

{\bf Lemma 6:} {\em Let $AB$ be a bi-symmetric composite of bi-symmetric models $A$ and $B$. 
Let $\B$ be a SPIN form on $\E(AB)$, and suppose that $\B$ {\em factors}, in the sense that, for all $x,x' \in X(A)$ and all $y, y' \in X(B)$, we have 
$\B(xy, x'y') = \B_1(x,x')\B_2(y,y')$ where $\B_1$ and $\B_2$ are normalized bilinear forms on $\E(A)$ and $\E(B)$, respectively. Then $\B_1$ and $\B_2$ are SPIN forms. If $A$ and $B$ are irreducible, 
then either (i) $\B$ is uniform, or (ii) $\B_1$ and $\B_2$ are orthogonalizing.}

{\em Proof:} That  $\B_1$ and $\B_2$ are both positive and symmetric is clear. To see that $\B_1$ is invariant, 
note that 
\[\B_1(a,b) = \B_1(a,b)\B_2(u_B,u_B) = \B(au_A, bu_B).\] Thus if $g \in G$, we have 
\begin{eqnarray*}
\B_1(ga,gb) = \B_1(ga, \overline{gb})\B_2(u_B, \overline{u}_B) & = & \B(ga u_B, \overline{gb} u_B)\\
&  = & \B((g,h)(a u_B) (g, h)(b, u_B))\\
&  = & \B(a,u_B,b,u_B) = \B_1(a,b)
\end{eqnarray*}
where $h \in H$ is arbitrary; similarly for $\B_2$. Now let $c, r$ be the parameters associated with $\B$, and let $c_1, r_1$ and $c_2, r_2$ be the parameters associated with $\B_1$ and $\B_2$, respectively. Let $x \perp y$ in $X(A)$ and $z \perp w \in X(B)$. As observed above\footnote{See the remarks following Definition 11}, it follows that $xz \perp yz$ and 
$x z \perp x w$, so we have 
\[c = \B(xz,yz) = \B_1(x,y)\B_2(z,z) = c_1 r_{2}^{2} \ \ \text{and also} \ \ c = \B(xz,yw) = \B_1(x,y)\B_2(z,w) = c_1 c_2.\]
Thus, $c_1 r_{1}^{2} = c_1 c_2$. If $B_1$ is not orthogonalizing, then $r_{1}^{2} = c_2$. Since $\B$ is irreducible,  $\B_2$ is uniform. But now the same reaoning, with 
the roles of $\B_1$ and $\B_2$ reversed, tells us that $c_1 c_2 = r_{1}^{2} c_2$, whence, as $c_2 \not = 0$, means that $c_1 = r_{1}^{2}$, whence, $\B_1$ is also uniform. But then $\B$ --- hence, $\omega$ --- is uniform as well. $\Box$ 

{\bf Definition 16:} Let $A$ and $B$ have conjugates $\overline{A}$ and $\overline{B}$. A state $\omega$ on $(AB)\overline{AB})$ is {\em factorable} iff 
iff there exist states $\omega_A$ on $AA'$ and $\omega_B$ on $BB'$ such that $\omega(xy\overline{xy}) = \omega_A(x\overline{x})\omega_B(y\overline{y})$. 

Applying Lemma 6 to the bilinear forms associated with the non-signaling states $\omega, \omega_A$ and $\omega_B$, we have 

{\bf Theorem 2:} {\em Let $A$ and $B$ be irreducible, and let $\omega$ be a factorable equivariant state on $(AB)(\overline{AB})$. Then either $\omega$ is the uniform state, or  $\omega_A$ 
and $\omega_B$ are orthogonalizing.}

Thus, if $A$ is irreducible, $B$ is a copy of $A$, and we can find a SPIN state on $(AB)(\overline{AB})$ making  $A\overline{A}$ and $B \overline{B}$ independent, we are guaranteed an orthogonalizing 
SPIN form. 

%{\em Remark:}  This has a little  of the flavor of the characterization of the standard normal density as the marginal 
%of a smooth radially symmetric bivariate density having independent marginals. \\
%I wonder if there's a connection.\\

\section{Monoidal Probabilistic Theories}

In the categorical approach to quantum foundations \cite{Abramsky-Coecke, Baez, Selinger}, it is usually assumed --- naturally enough --- that a 
physical theory is a symmetric monoidal category $\C$, in which objects represent physical systems, morphisms represent 
physical processes, and the tensor product represents the physical composition of systems. A stronger, and perhaps more mysterious, assumption is that $\C$ be {\em dagger-} monoidal, i.e, that it carry an involution compatible with the 
monoidal structure. In this section, I consider a symmetric monoidal category $\C$ of bi-symmetric probabilistic models, and consider the associated ``linearized" category $\E({\cal C})$ consisting of the linear hulls of the models in $\C$. The main result is that, if $\E({\cal C})$ is consistent with the existence of reasonable dagger-monoidal structure (the adjective ``reasonable" being spelled out in Definitions 19 and 20 below), then there exists a factorable SPIN form on $\E(A)$ for each model $A \in \C$; hence, irreducible models in $\C$ carry orthogonalizing SPIN inner products. Add the requirement that models in $\C$ are sharp and state-complete, and all models in $\C$ are self-dual.

{\bf 4.1 Monoidal categories of probabilistic models}

Henceforth, $\C$ will denote a category of models, with morphisms as defined in section 1.4.  It is reasonable to require --- and I shall require --- a bit more, namely, that every symmetry $g \in G_A$ is in fact a morphism in $\C$, i.e., $G_A \leq \C(A,A)$. 

A {\em symmetric monoidal structure} on a category $\C$ is a bifunctor $\otimes : \C \times \C \rightarrow \C$, plus 
a designated {\em unit object} $I \in \C$, and natural isomorphisms 
\[\alpha_{A,B,C} : A \otimes (B \otimes C) \simeq (A \otimes B) \otimes C, \ \sigma_{A,B} : A \otimes B \simeq B \otimes A,\]
\[ \lambda_A : I \otimes A \simeq A, \ \text{and} \ \rho_A : A \otimes I \simeq A\]
for all $A, B, C \in \C$. These isomorphisms are also required to satisfy various coherence conditions, e.g., 
that  $\lambda_{A} \circ \sigma_{A,I} = \rho_{A}$. See \cite{Maclane} for details. A {\em symmetric monoidal category} is a category equipped with such a structure. 
A {\em dagger} on a SMC $\C$ is an endo-functor $\dagger : \C \rightarrow \C$ such that, for 
all objects $A \in \C$, $A^{\dagger} = A$, $\sigma_{A,B}^{\dagger} = \sigma_{B,A}$, and, for all morphisms $\phi, \psi$ in $\C$,
$\phi^{\dagger\dagger} = \phi$  and $(\phi \otimes \psi)^{\dagger} = \phi^{\dagger} \otimes \psi^{\dagger}$. 
A {\em dagger-monoidal} category is a SMC equipped with a dagger.  

{\bf Definition 18 (Monoidal probabilistic Theories):} A {\em monoidal probabilistic theory} is a symmetric monoidal category of probabilistic models, such that 
\begin{itemize} 
\item[(i)] $G(A) \leq \C(A,A)$ for every $A \in \C$,
\item[(ii)] for every $A, B \in \C$, the monoidal product $A \otimes B$ defines a non-signaling composite of $A$ and $B$, in the sense of Definition 11 
\item[(iii)] the morphism $A \times B \rightarrow A \otimes B$ sending $x \in X(A), y \in X(B)$ to $xy \in X(A \otimes B)$, is a morphism in $\C$.  
\end{itemize}

For each model $A \in \C$, we have the corresponding linear hull, the order-unit space $\E(A)$.  Now, every morphism 
$\phi \in \C(A,B)$ defines an affine mapping $\phi^{\ast} : \Omega(B) \rightarrow \Omega(A)$, given by $\phi^{\ast}(\beta)(x) = \beta(\phi(x))$. Pulling back again, we have a linear 
mapping $\phi^{\ast \ast} : \Aff(\Omega(A)) \rightarrow \Aff(\Omega(B))$, which evidently takes $\E(A)$ to $\E(B)$. Thus, 
$A \mapsto \E(A)$ is the object part of a covariant functor from $\C$ to the category of ordered linear spaces and  positive linear mappings, taking $\phi \in \C(A,B)$ to the 
coresponding positive linear mapping $\phi : \E(A) \rightarrow \E(B)$ 
where $\phi(a)(\beta) = a(\beta \circ \phi)$ for all $a \in \E(A)$ and $\beta \in \Omega(B)$. %[But: overload notation...]

%At this point, I want to assume that the category $\C$, or, better, its ``linearized" image category 
%$\E(\C)$, is compatible with a dagger-monoidal structure. Roughly, this means that it is possible to enlarge the category $\E(\C)$, by adding morphisms, so as to obtain a dagger-monoidal category of %order-unit spaces. Actually, I want to be a bit more restrictive than this: I want the dagger on ${\cal E}$ to reduce to inversion when applied to an element of $G_A$, $A \in \C$. Let me make this %more formal:

It is easy to check that the unit object for a monoidal probabilistic theory will necessarily be the trivial model $T = (\{1\}, \{\{1\}\}, \{1\}, \{e\})$ having one outcome, one test, one state, and one symmetry. It is 
an annoying fact that there are {\em no} morphisms between $T$ and any non-trivial model. 
%In particular, then, 
%we can't interpret outcomes $x \in X(A)$ as morphisms $x : I \rightarrow A$ or $A \rightarrow I$ [explain why we'd want to...].  
The linearized category $\E(\C)$ will inherit the same defect. Thus, we'd like to extend the 
set of morphisms in the latter -- at a minimum, we'd like to allow arbitrary linear mappings 
$\E(T) \simeq {\Bbb  R} \rightarrow \E(A)$ --- representing elements of $\E(A)$ --- as well as some linear mappings $\E(A) \rightarrow {\Bbb  R}$, representing states, to count as morphisms. This suggests the following 

{\bf Definition 19 (Representations):} A {\em representation} of a probabilistic theory $\C$ is a functor $\pi : \C \rightarrow {\cal E}$ where ${\cal E}$ is a 
\ category of order-unit spaces, such that
\begin{itemize} 
\item[(i)] for all $V, W \in {\cal E}$, ${\cal E}(V,W)$ is a space of linear mappings, ordered by a cone ${\cal E}_+(V,W)$ of positive linear mappings; 
\item[(ii)] $\pi(A) = \E(A)$ for every $A \in \C$, and $\pi(\phi) = \phi$ for every $\phi \in \C(A,B)$, 
\item[(iii)] ${\cal E}({\Bbb  R}, \E(A)) \simeq \E(A)$, for all $A \in {\cal E}$. 
\end{itemize} 
A representation is {\em self-dual}, resp. HSD, iff  $\pi(A)$ is self-dual, respectively HSD, for every model $A \in \C$. 

If $\C$ is a monoidal probabilistic theory, we can ask that ${\cal E}$ also be symmetric-monoidal, and that $\pi$ be a monoidal functor, i.e., $\pi(AB) = \pi(A) \otimes \pi(B)$ (at least up to a canonical isomorphism). In this case, we shall say that $\pi$ is a {\em monoidal representation} of $\C$. 
%By a {\em dagger-monoidal} representation, I mean a monoidal representation 
%$\pi : \C \rightarrow {\cal E}$ where ${\cal E}$ is dagger-monoidal. 

%and such that the monoidal product $A \otimes B$ of $A, B \in \C$ is a non-signaling composite of $A$ and $B$. One would also {\em like} to be able to insist that, for models $A, B \in \C$, 
%the canonical mapping $x,y \mapsto xy$ defining $A \otimes B$ as a composite, is precisely the morphism $ 

%[WANT to say $xy = x \otimes y$. But $x, y$ are not (yet) morphisms. Need to linearize!]

%It would be very helpful to be able to treat outcomes $x \in X(A)$ as arrows $x : I \rightarrow A$ 
%(I) Image category linear, but still not have enough structure. 
%
%(II) Representation: $\E(A) \leq {\cal E}$ with same objects, but more morphisms: monoidal, dagger, etc....

%Next: daggers, fact that $\eta_A$ induces a dagger. (But maybe this goes in the next section?) 

{\bf 4.1 Dagger-Monoidal Representations} 

A basic assumption in the categorical approach to finite-dimensional quantum theory \cite{Abramsky-Coecke, Baez, Selinger} is that the category of physical systems and processes should be, not just a symmetric monoidal, but a {\em dagger}-symmetric monoidal category. Roughly, the monoidal product $A, B \mapsto A \otimes B$ is understood to capture the idea of a composite of two non-interacting (but possibly entangled) systems; the meaning of the dagger is a perhaps a bit more mysterious, but is suggestive of an operation of time-reversal. %... [ABOVE?]

%{\bf Definition:} $\C$ is {\em self-dual} iff, for every $A \in \C$, the cone $\E(A)_+$ is homogeneous and %self-dual. 

%[Linearizing $\C$...]

%{\em {\bf NOTE} that we do need the $G_A$'s to be subgroups of the unitary groups $U(A) = \{ g \in \C(A,A) | g^{\ast} = %g^{-1}\}$. This may require some further justification!! (Another way to say this is that the dagger must extend the operation %of inversion on the groups.)}

{\bf Definition 20 (Dagger-monoidal representations):} A {\em dagger-monoidal representation} of a monoidal category $\C$ of probabilistic models is a monoidal 
representation $\pi : \C \rightarrow {\cal E}$ where 
\begin{itemize} 
\item[(i)]  $\cal E$ is $\dagger$-monoidal, with $I \simeq {\Bbb  R}$,
\item[(ii)] $u_{A}^{\dagger} \circ u_{A} = 1$ for all $A \in \C$, and 
\item[(iii)] $\pi(g^{-1}) = \pi(g)^{\dagger}$ for all $g \in G(A)$, $A \in \C$. 
\end{itemize} 
I'll say that $\C$ is {\em dagger-monoidal} iff it has a dagger-monoidal representation. 
% and {\em HSD} iff it 
%has a homogeneous, self-dual representation. 

%[Note that, by (i), have [or is this another hypothesis?] that $x \otimes y =: xy$ is the canonical injection from 
%$X_A \times X_B$ into $X_{A \otimes B}$. Does this IMPLY non-signaling??]

%{\bf Standing assumption:} {\em Henceforth, $\C$ is a symmetric monoidal category of fully-symmetric probabilistic %models, admitting a dagger-monoidal representation.}

Subject to  assumptions (i) and (ii), there exists, for each $A \in \C$, a canonical 
$G(A)$-invariant, positive, symmetric bilinear form on $\E(A)$ given by 
\begin{equation}\langle a, b \rangle := a \circ b^{\ast}\end{equation}
with $a, b \in \E(A) \simeq {\cal E}(I, \E(A))$.\footnote{In general, this is not an inner product; the angle-bracket notation is, however, standard in this context.} 
%This is even normalized, i.e., %$\langle u, u \rangle = u \circ u^{\dagger} = 1$. 
Now, just by virtue of the monoidality of $\cal E$, this bilinear form {\em factors}, 
in the sense that 
\[\langle a \otimes b, c \otimes d \rangle = \langle a, c \rangle \langle b, d \rangle.\]
This at once yields %[Should THIS be Theorem 1?]

{\bf Theorem 3:} {\em  Suppose $\C$ is $2$-symmetric, and admits a dagger-monoidal representation. Then the 
canonical form (\theequation) is orthogonalizing on every irreducible 
system $A \in \C$.} % of rank $\geq 2$.} 

The proof is virtually identical to that of Theorem 2. 

%{\em Proof:} We can suppose that the rank of $A$ is at least $2$ (the rank-one case being trivial). Let 
%$r$ and $c$ be the parameters associated with the canonical form on $\E_A$, and $c_2, r_2$, the 
%corresponding parameters for $\E_{A \otimes A}$.  
%If $x,y \in X_{A}$ with  $x \perp y$, then in $X_{A \otimes A}$, we have 
%\[x \otimes x \perp x \otimes y \ \text{and} \ x \otimes y \perp y \otimes x.\]
%As the canonical form is bi-functorial, we therefore have 
%\[c_2 = \langle x \otimes x, x \otimes y \rangle = \langle x, x \rangle \langle x, y \rangle = r^2 c\]
%and also 
%\[c_2 = \langle x \otimes y, y \otimes x \rangle =  \langle x, y \rangle \langle x,y \rangle = c^2.\]
%Hence, $r^2 c = c^2$, whence, either $c = 0$ or $c = r^2$. In the latter case, $r^2 = 1/n^2$, and $\dim(\E) = 1$, whence, 
%the rank of $\A$ is $1$. $\Box$%

Combining this with Lemma 2 and Corollary 2, we have 

{\bf Corollary 4:} {\em Let $\C$ be as in Theorem 3, and let $A \in \C$ be state-complete, sharp and irreducible. 
Then the canonical bilinear form (\theequation) is an inner product, with respect to which $\E(A)$ is self-dual.  }\\

I'll call $\C$ {\em $\dagger$-Self Dual} ($\dagger$-SD) iff there exists a $\dagger$-monoidal representation $\pi : \C \rightarrow {\cal E}$ where 
each $\pi(A)$ is self-dual with respect to (\theequation) --- meaning, in particular, that this blinear form is an inner product for 
every $A \in \C$. Corollary 2 tells us that if $\C$ has a dagger-monoidal representation {\em and} every $A \in \C$ is state-closed, sharp, and irreducible, then 
$\C$ is dagger-SD.

{\em Remark:} Suppose that every object $A \in \C$ has a conjugate $(\bar{A}, \eta_A, \gamma_A)$ with $\bar{\bar{A}} = A$, in the sense that 
$\bar{\bar{A}} = (A,\eta_A \circ \sigma, \gamma_{A}^{-1})$ (here $\sigma : \bar{A} \times A \rightarrow A \times \bar{A}$ is the obvious swap mapping). If the correlators $\eta_A$ are 
symmetric, in the sense that $\eta_{A}(a,\bar{b}) = \eta_{A}(b, \bar{A})$ for all $a, b \in \E(A)$, then one can construct a dagger on $\E(A)$ as follows: if $\phi \in \C(A,B)$, set 
\[\phi^{\dagger} = \tau_{A}^{-1} \circ \phi^{\ast} \circ \tau_B\]
where $\tau_A : \E(A) \rightarrow \E(A)^{\ast}$ is the mapping given by $\tau_{A}(a) = \widehat{\eta}^{\ast}(\gamma_{A}(a))$, i.e., $\tau_{A}(a)(b) = \eta_{A}(b, \gamma_{A}(a))$. With this 
definition of $\phi^{\dagger}$, one has $\langle \phi(a),b \rangle = \langle a, \phi^{\dagger}(b) \rangle$ where $\langle a, b \rangle := \eta_{A}(a, \gamma_{A}(b))$, as above. 
(It would be interesting to know how the existence of a canonical, involutive operation $A \mapsto \bar{A}$ of conjugation on a monoidal probabilistic theory comes to making the 
category dagger-compact.)
%[Does $\langle, \rangle$ then agree with the one given by the dagger? Yes --- but check: $a, b \in \C(I,A) \simeq A$ implies $a^{\dagger} \circ b = \langle a, b \rangle$. ]

%{\em Remark:} Another way to say this is that, in any dagger-monoidal category of order-unit spaces, endowed with a privileged %family of observables, fully-symmetric under a compact group of unitaries, irreducible spaces are self-dual. It follows that %such a category is also $\dagger$-compact (\cite{BDW}, Corollary 28). \\

%%Cats of JAs: all are direct sums of irreps. So, no need for IC??

%%\newpage
\section{Image-closure}

Thus far, our results mainly concern irreducible systems. One way of extending them to possibly reducible systems involves a condition I'll call {\em image-closure}.  

%[NOTE: In LTHSD, we note the fact that, for an {\em irreducible} homogeneous, self-dual cone $A_+$, the space $u^{\perp}$ complementry to the order unit (with respect to self-dualizing inner product) is irreducible  as a $K_A$-module \cite{Vinberg2}. So we can observe that if we pass from $G_A$ to $K_u$, this gives us a bigger  but still (?) FS test space, AND lets us substitute irreducibility (here, indecomposability!] for OUR weaker notion of irreducibility, which is to say --- we can omit this section! [CHECK!] Actually, that's not quite right. We can replace parts of this section --- Hm. Need to say that (i) every cone's a convex sum of irred. cones; (ii) if every summand is in the category (this seems not so unreasonable) then we're d one... Eh! NOT SO FAST: the assumption would have to be that the SUM is already SD, which would be circular here.]

{\bf Definition 21 (Image of a model):} A morphism $(\phi,\psi) : A \rightarrow B$ is {\em surjective} iff $\phi(X) = Y$, $ \subseteq \phi(\A)$, $H = \psi(G)$, and $\Omega(B) = \{ \beta \in \Omega(X(B), \A(B)) | \phi^{\ast}(\beta) \in \Omega\}.$ In this case, we call $B$ the {\em image} of $A$ under $(\phi,\psi)$, writing $B = \phi(A)$. 
%$\psi : G \rightarrow H$ is surjective.  [Do we need this last? Would it not be better to allow maximal state space...?]

Notice that the image $B = \phi(A)$ can be {\em simulated} by $A$, as follows. To prepare $B$ in state $\beta$, prepare the $A$ in the state $\phi^{\ast}(\beta)$. To measure $F = \phi(E)$, 
measure $E$ on $A$, and, upon obtaining outcome $x \in E$, record $\phi(x)$ as the outcome of $F$. To implement a symmetry $h \in H$, implement any corresponding symmetry 
$g \in \psi^{-1}(h) \subseteq G$. Operationally, it is reasonable (so long as we can prepare arbitrary states) to take $\phi(A)$ as a legitimate physical model whenever $A$ is. 
 
{\bf Definition 22 (Image-closure):} Call $\C$ {\em image-closed} iff, for any model $A \in \C$ and any surjective morphism $\Phi = (\phi,\psi) : A \rightarrow B$, 
(i) the model $B$ belongs to $\C$, and (ii) $\Phi \in \C(A,B)$. 

{\em Remark:} The image of a $2$-symmetric model is $2$-symmetric. 

%Operationally, it is reasonable to suppose that a probabilistic theory should be image-closed. For if $(\phi,\psi) : (X,\A,\Omega,G) \rightarrow (Y,,\Gamma,H)$ is surjective, then 
%any test $F \in $ can be executed (or perhaps we should say, simulated) simply by choosing a test $E \in \A$ with $\phi(E) = F$, and, upon securing an outcome $x \in E$,  recording $\phi(x)$ rather than $x$. Similarly, all symmetries in $H$ can be implemented by appropriate symmetries in $G$. Finally, if $\beta \in \Gamma$, then the very definition of a morphism of models tells us that 
%$\beta(\phi(x)) = \alpha(x)$ for $\alpha = \phi^{\ast}(\beta)$, so that, to prepare $\beta$, we simply prepare $\alpha \in \Omega$. (Note here that $\beta(\phi(x)) := \alpha(x)$ is well-defined.) In effect, we are (i) coarse-graining $(X,\A)$, and (ii) preparing a restricted set of states. As long as we believe that we can prepare arbitrary states in $\Omega$, and so long as we are permitted to restrict our attention to any closed, convex set of states we like, this should yield an allowed model.\footnote{To be sure, this suggests that $\C$ ought to be closed under the 
%replacement, for any model $A \in \C$, of the state space $\Omega_A$ by any smaller (say, closed) convex set. However, the condition of image closure is weaker --- and thus, surely at least as reasonable --- as this.}%[THIS is still a bit lame. Either arbitrary restricted state spaces are allowed, or not. If not, then neither are 
%arbitrary images, as not all states in $\Omega$ will be compatible with the structure of $\phi(\A)$. ]

It would be rather embarrasing, at this point, if the category $\C_{QM}$ of quantum models were not image-closed. In fact, however, a quantum model has no non-trivial images at all. 

{\bf Definition 23 (Incompressible Models):} A model $A$ is {\em incompressible} iff, for all models $B$, any surjective homomorphism $\phi : A \rightarrow B$ 
is either an isomorphism, or is trivial in the sense that $X_B$ is a single point.

{\bf Lemma 7:} {\em Every quantum model is incompressible.}

{\em Proof:}  see this, notice first that if $\phi : A \rightarrow B$ is a surjective morphism of models, with $G$ acting transitively on $X(A)$, then $X(B)$ is a transitive $G$-set. Hence, $X(B) \simeq G(A)/G(A)_y$, where $G(A)_y$ is the stabilizer of any $y \in X(B)$ under the action of $G(A)$ on $X(B)$. Since $\phi$ is equivariant, $G(A)_x \leq G(A)_y$  for any $x \in X(A)$ with $\phi(x) = y$. Now suppose that $A$ is quantum, i.e., $A = A(\H)$ for 
a complex $n$-dimensional Hilbert space $\H$. Then $X(A) = X(\H)$, the set of rank-one projections on $\H$, and $G(A) = U(\H) \simeq U(n)$, with $G(A)_x \simeq U(n-1)$. Now, $U(n-1)$ is a maximal proper subgroup of $U(n)$\footnote{Thanks to David Feldman for pointing this out}. Hence, as $G(A)_x \leq G(A)_y$, we have either $G(A)_y = G(A)_x$ or $G(A)_y = G(A)$.  In the former case, $\phi$ is a bijection, in the latter, $X(B)$ is a point. Thus, a non-trivial surjective image of a quantum model $A(\H) = (X(\H),\A(\H),\Omega(\H),U(n))$ has the form $(X(\H), \A(\H), \Gamma, U(n))$, where $\Gamma$ is the set of 
all states on $(X(\H),\A(\H))$ of the form $\rho \circ \phi$ where $\rho$ is (the state associated with) a density operator on $\H$ and $\phi$ is a symmetry of $(X(\H), \A(\H))$. By Wigner's Theorem, $\phi$ has the form 
$\phi(x) = V^{-1} x V$ where $V : \H \rightarrow \H$ is either  
 unitary or anti-unitary. We have, for every unit vector $x \in X(\H)$, 
\[(\rho \circ \phi)(x) = \Tr(\rho \phi(x)) = \Tr(\rho V^{-1} x V) = \Tr(V \rho V^{-1} x) \]
That is, the state $\rho \circ \phi$ is the state on $(X(\H),\A(\H))$ associated with the linear operator 
$V \circ \rho \circ V^{-1}$ --- which is a linear, and thus a density operator, regardless of whether 
$V$ is linear or anti-unitary. Thus, $\Gamma = \Omega(\H)$.
\footnote{If $\dim(\H) > 2$, Gleason's Theorem 
gives us $\rho \circ \phi \in \Omega$ even more trivially. The preceding argument also works if $\dim(\H) = 2$.} 
$\Box$

{\em Remark:} This argument shows that any model $A$ such that (i) $X(A)$ transitive under $G(A)$, 
(ii) the stabilizer $G(A)_x$ of an outcome $x \in X(A)$ is a maximal subgroup of $G(A)$, and (iii) $\Omega(A)$ invariant under symmetries of $(X(A),\A(A))$, is incompressible. 

Our goal now is to prove the following:

{\bf Theorem 4:} {\em Suppose $\C$ is an image-closed category of bi-symmetric models, having a dagger-monoidal 
representation. 
% and $\E(A)$ carries an orthogonalizing SPIN inner product for every irreducible system $A \in \C$. 
Then $\E(A)$ carries an orthogonalizing SPIN inner product for $A \in \C$.} 
%$u_{A}^{\perp} \leq \E_A$ is an irrep of $G_A$.} 
%[NOTE: need to leverage Lemma ... to avoid explicit sharpness axiom...]

Let $(X,\A,\Omega,G)$ be a fully symmetric model. Let $\langle, \rangle$ be a SPIN inner product on $\E$, e.g., 
the one arising from group averaging. Suppose $\M \leq u^{\perp}$ is a $G$-invariant subspace of $u^{\perp}$. Let $\p : \E \rightarrow \M$ be the corresponding projection operator (defined w.r.t. the standard inner product). For every $x \in X$, set 
\[x_1 := \p(x) + u/n\]
Then $\sum_{x \in E} x_1 = \p(u) + u = u$ 
(with $\p(u) = 0$ since $\M \leq u^{\perp}$).  Let $X_1 = \{ x_1 | x \in X\}$; for each $E \in \A$, set $E_1  = \{ x_1 | x \in E\}$, and let $\A_1 = \{ E_1 | E \in \A\}$. Then $(X_1,\A_1)$ is a fully symmetric $G$-test space. 
%and $x \mapsto x_1$ is  
Since $X$ spans $\E$, $X_1$ spans $\E_1 := \M \oplus \langle u \rangle$. Let $\E_{1 +}$ denote the cone in 
$\E_1$ consisting of non-negative linear combinations of elements of $X_1$. 
%Let $\E_1 \leq \E$ denote the 
%span in $\E$ of $X_1$, ordered by the cone 
%consisting of non-negative linear combinations of elements of $X_1$.  %[Supply proof?]

{\bf Lemma 8:} {\em There exists a $G$-invariant, separating set $\Omega_1$ of states on $(X_1,\A_1)$ such that (i) $A_1 := (X_1,\A_1,\Omega_1, G)$ is a bi-symmetric model (in particular, $G$ acts transitively on the extreme points of $\Omega_1$), and (ii) the pair 
$(\phi,\id_G)$, with $\phi : X \rightarrow X_1$ given by $\phi(x) = x_1$, is a morphism of models. }

{\em Proof:} Let $v \in \E^{+}$ represent a pure state, i.e., an extreme point of $\Omega$. Set $v_1 = \p(v) + u$, 
and note that \[\langle v_1, u \rangle = \langle \p(u),u \rangle + \langle u, u \rangle = 1\]
and, for all $x \in X$, 
\[\langle v_1, x_1 \rangle = \langle \p(v) + u, \p(x) + u/n \rangle = \langle \p(v),\p(x) \rangle + 1/n = \langle v_1, x \rangle.\]
There is no gaurantee that this last will be positive for all $x \in X$; however, we can choose $\epsilon > 0$ 
so that 
\[v_{\epsilon} := \epsilon v_1 + (1 - \epsilon)u\]
belongs to $\E^{+}$ -- and hence, to $\E_{1}^{+}$ --- since $u$ lies in the interior of $\E^{+}$. 
%(Here, $\E_{1}^{+}$ is the 
%internal dual of $\E_{1,+}$, the positive span of $X_1$ in $\Aff(\Omega)$.) 
Now let 
\[\Omega_{1} := \overline{\text{co}}(G v_\epsilon)\]
This is clearly a closed, convex, $G$-invariant set of states on $(X_1,\A_{1})$. We must show it is separating. 
Suppose 
\begin{equation} 
\langle g v_{\epsilon}, x_1 \rangle = \langle g v_{\epsilon}, y_p \rangle \end{equation}
for all $g \in G$. We have 
\begin{eqnarray*}
\langle g v_{\epsilon}, x_{1} \rangle & = & \langle \epsilon v_1 + (1 - \epsilon)u, g^{-1}x_1 \rangle \\
& = & \langle \epsilon v_1 + (1 - \epsilon) u, g^{-1} \p x + u/n \rangle \\
& = & \epsilon \left ( \langle v_1, g^{-1}\p x \rangle + \langle v_1, u/n\right ) + 
(1 - \epsilon) \left ( \langle u, g^{-1}\p x \rangle + \langle u, u/n \right)\\
& = & \epsilon \langle v_1, g^{-1} \p x \rangle + \epsilon 1/n + (1 - \epsilon) 1/n \\
& = & \epsilon \langle v_1, g^{-1} \p x \rangle + 1/n.
\end{eqnarray*}
Similarly, $\langle v_{\epsilon}, y_1 \rangle = \epsilon \langle v_1, g^{-1} \p y \rangle + 1/n$. Thus, 
(\theequation) implies $\langle v_1, g^{-1} \p x \rangle = \langle v_1, g^{-1}\p x \rangle $, 
whence, $\langle \p(v) + u, g^{-1}\p(x) \rangle = \langle \p(v) + u, g^{-1}\p(y) \rangle$, whence, 
$\langle \p(v), g^{-1}\p(x) \rangle = \langle \p(v), g^{-1}\p(x) \rangle$. But this last is 
\[\langle gv, \p x \rangle = \langle gv,\p y \rangle \]
for all $g \in G$. Since $\{ g v | g \in G\}$ is the full set of extreme points of $\Omega$, it is separating 
for $\E$. It follows that $\p(x) = \p(y)$, i.e, $x_1 = y_1$. $\Box$ 

%Notice that $\Omega_1 - u/n$ contains an open neighborhood of the origin in $\M$, hence, spans $\M$. Thus, 
If $\M$ is a minimal proper $G$-invariant subspace of $u^{\perp}$, the model $A_1$ is irreducible.  
Hence, if $A_1$ supports an orthogonalizing positive symmetric invariant bilinear form, 
then (by Corollary 2) this form is the standard SPIN inner product $\langle ~,~\rangle_1$ on $\E_1$. 
%(Note that the standard inner product on $\E_1$ is not (or not obviously) the standard inner product on $\E$, restricted to %$\E_1$.)

{\em Remark:}  The foregoing proof shows that if $A$ is incompressible, then $u^{\perp}$ is irreducible in $\E(A)$.  
Thus, we can dispense with image-closure, if we are willing to focuss our attention on incompressible models: 

{\bf Theorem 4b:} {\em Let $\C$ be a dagger-monoidal category of $2$-symmetric probabilistic models. Then 
for every incompressible model $A \in \C$, $\E(A)$ hosts an orthogonalizing SPIN inner product. If 
$A$ is also state-closed and sharp, then $\E(A)_+$ is self-dual.}

%[But better double-check the point about states... Hm! Just need to observe that, since image under $\pi$ DOES have a separating set of states, which pull back to states on the initial model, we can always extend to a suitable $\Gamma$...]

Returning now to a general situation, let $u^{\perp} = \M_1 \oplus\cdots \oplus \M_k$ with each $\M_j$ an irreducible invariant subspace for $G$. Let  $\p_1,....,\p_{k}$ be the corresponding projections, and, for each $x \in X$, let $x_j = \p_j(x) + u/n$, $j = 1,...,n$.  Lemma 8 gives us, for each $j$, a bi-symmetric 
model $(X_j,\A_j,\Omega_j,G)$, and, with this, a space $\E_j = \M_j \oplus \langle u \rangle$ (ordered by 
the cone spanned by $X_j$). Finally, since each $A_j$ is irreducible, Corlollary 2 gives us a standard 
(maximal) SPIN inner product $\langle , \rangle_{j}$ on $\E_j$ 

{\bf Lemma 9:} {\em If $\langle , \rangle_j$ is orthogonalizing for each $j$, then there exists an orthogonalizing inner product 
on $\E$.} 

{\em Proof:} With notation as above, let 
\[\langle a, b \rangle_{\ast} = \sum_{j=1}^{k} \langle \p_j(a),\p_j(b) \rangle_{j} + k\langle a,u \rangle \langle b,u \rangle.\]
This is clearly bilinear, invariant and symmetric. Indeed, since each $\langle , \rangle_j$ is an inner product, so is 
$\langle , \rangle_{\ast}$. To see that it is positive on $\E_+$, note that for every $x \in X \subseteq \E$, 
$x = (\sum_{j} \p_i x)  + u/n$, so, for $x, y \in X$, we have 
\begin{eqnarray*} 
\langle x, y \rangle_{\ast} & = &  \sum_{j} \langle x_j, y_j \rangle + k /n^2 \\
& = & \sum_{j} \langle \p_j(x) + u/n, \p_j(y) + u/n \rangle_j \\
& = & \sum_{j} \langle x_j, y_j \rangle_j \geq 0.
\end{eqnarray*} 
Since $X$ spans $\E_{+}(A)$, $\langle ~,~ \rangle_{\ast}$ is positive. 
The same computation shows that if $x \perp y$, so that $x_j \perp y_j$ for each $j$, then, as $\langle x_j, y_j \rangle_j = 0$ by hypothesis, $\langle x, y \rangle_{\ast} = 0$. $\Box$ 

%[NOTE: THIS SEEMS TO DEPEND ON $\E_+ = \Span_+(X)$!. Can we fix? (ALSO: what happened to $+1/n$ in last line?]

%{\bf Corollary 2:} {\em Under the assumptions of Lemma 4, $u^{\perp}$ is irreducible (that is, all but one $P_j$ is $0$.}

{\em Proof of Theorem 4:} Let $A = (X,\A,\Omega,G)$ be a model in $\C$, and proceed as above to construct 
models $A_j = (X_j,\A_j,\Omega_j,G_j)$ corresponding to the irreducible components of $u^{\perp}$ in $\E = \E(A)$.  
By Lemma 8, $A_j$ is the image of $A$ under a surjective homomorphism. Since $\C$ is image-closed, each $A_j$ lies in $\C$. Since $\C$ is also dagger-monoidal, we have a canonical invariant bilinear form (\theequation) on each $\E(A)$, $A \in \C$, and this is orthogonalizing. Since $A_j$ is irreducible, Corllary 2 tells us that 
this canonical form on $\E(A_j) = \E_j$ must coincide with standard form $\langle , \rangle_j$ 
for all $j = 1,...,n$, whence, the SPIN  inner product $\langle ~,~ \rangle_{\ast}$ of Lemma 9 is orthogonalizing. Theorem 4 now follows from  $\Box$ 

%[HERE, WE NEED TO SHOW THAT WE CAN TRADE $\Omega_1$ for $\Omega_{1}'$ generated by $\delta_1 := \langle \pi x_o |_1$ for 
%some $x_o \in X$. BUT, since $\pi^{\ast}(\delta_1)(a) = \langle \pi x_o, \pi a \rangle_1$ and $\pi$ is positive, 
%we have this $\geq 0$ on $\E_+$, i.e., in $\E^{\ast}_+$ --- meaning it's a multiple of something in $\Omega$, yes? 
%CHECK...] 

This gives us %the %Theorem B from the introduction, which here I'll restate as a 

{\bf Corollary 5:} {\em If $\C$ is an image-closed monoidal category of $2$-symmetric models, admitting a $\dagger$-monoidal 
representation, then  every state-closed, sharp model $A \in \C$ is self-dual.} \\

%NOTE: The cat. or 2-symm models IS image closed (by virtually by definition of surjective image). Suppose RAR is %terminal, i.e., has no non-trivial hom. images (should we say rather than $A$ is {\em incompressible?}). If $A$'s %sharp, $\E(A)_+$ is self-dual. \\
%%%\newpage

{\bf 6. Homogeneity} %and the Existence of a Dagger}\footnote{Reorganize to stress Iso-Dil.}

Let $\C$ be an image-closed category of $2$-symmetric probabilistic models. We've seen that if every model in $\C$ has a conjugate, or if $\C$ has a $\dagger$-monoidal representation, then for every 
state-closed, sharp model $A \in \C$, the cone $\E(A)_+$ is self-dual. If this cone is also homogeneous, then the Koecher-Vinberg Theorem tells us that $\E(A)_+$ is isomorphic to the cone of squares of a formally real Jordan algebra. 

%If we assume that composites in $\C$ are non-signaling --- not an unreasonable assumption! --- then 
There are several ways in which to motivate the homogeneity of $\E_+(A)$, earlier 
explored in \cite{Wilce11b} and \cite{BGW}. Before discussing these, let me mention one very direct interpretations of 
homogeneity. If we allow that all order-automorphisms $\phi$ of $\E^{+} \simeq \V(\Omega)$ with $u(\phi(\alpha)) \leq u(\alpha)$ represent legitimate physical processes, then homogeneity simply requires that it be possible to prepare any state in 
the interior of the cone, with non-zero probability, by applying a reversible physical process to the maximally mixed state. 
The main objection to simply taking this as a postulate is probably just that the use of the adjective ``interior" here seems unaesthetic. 
(Then again, we seldom scruple to accord special axiomatic privileges to pure states.)

{\bf 6.1 Self-Steering and Iso-Dilation} 

In \cite{BGW} it is shown that homogeneity of the state cone follows from the assumption that every $A \in \C$ is ''self-steering": 

{\bf Definition 24 (Self-Steering)):} A system $A$ has the {\em Self-Steering property} iff every state $\alpha \in \Omega(A)$ arises as the marginal of some bipartite state $\omega \in \Omega(A \otimes A)$ that is {\em steering}, in the sense that, for every convex 
decomposition $\sum_i t_i \alpha_i = \alpha$ of $\alpha$ as the average of an ensemble of other states, 
there exists an observable ${\cal E} = \{a_i\}$ on $\E(A)$ with $\omega(a_i, \cdot) = t_i \alpha_i$ for each $i$. 
%[Benefits of 

A less vivid, but mathematically simpler, assumption, also discussed in \cite{BGW}, is that every state in the interior of the state space, arise as the marginal of --- or, in other language, can be {\em dilated to} --- a bipartite {\em isomorphism state}, that is, a state $\omega$ whose conditioning map, $\widehat{\omega}$, is an order-isomorphism $\E^{\ast} \simeq \E$.  
We might call this the {\em Iso-Dilation} condition. To see that this implies homogeneity, simply note that if $\alpha$ and $\beta$ are any two interior states (not necessarily normalized), then by assumption there exist bipartite states $\omega_1$ and $\omega_2$ with $\widehat{\omega}_1(u) = \alpha$ and $\widehat{\omega}_2(u) = \beta$, whence, $(\widehat{\omega}_2 \circ \widehat{\omega}_{1}^{-1})(\alpha) = \beta$. Of course, there is still the (dubious?) aesthetic  objection regarding the interior states. 

That Self-Steering implies the homogeneity of the state cone is a consequence of the fact that any steering state 
on $A \otimes A$ having a marginal lying in the interior of the state cone, must be an isomorphism state. {\em A priori}, then, Iso-Dilation is the weaker condition. When $V(A)_{+}$ is irreducible, isomorphism states are pure, so this is a relative of the ``purification postulate" of \cite{CDP}. \\

%Here are three:
%[Don't we need to assume that marginals are states in $\C$?]
%[BUT: These need LT and NS?!]

{\bf 6.2 Full symmetry, correlation and filtering} 

Suppose $A \in \C$ is sharp and {\em fully} symmetric, rather than only $2$-symmetric. Then we can use the ``correlation" and ``filtering" axioms from \cite{Wilce11b} to secure the homogeneity of $\E(A)_+$. Recall that a bipartite state $\omega$ on 
a composite $AB$ {\em correlates} tests $E \in \A(A), F \in \A(B)$ iff there is a bijection $f : E \rightarrow F$ such that 
for all $x, y \in E \times F$ with $y \not = f(x)$, $\omega(x,y) = 0$. In other words, on $E \times F$, $\omega$ is supported on the graph of $f$. In this situation, I'll say that $\omega$ correlates $E$ and $F$ {\em along} the bijection $f$.

{\bf Definition 25 (Correlation Condition):} A model $A$ satisfies the {\em correlation condition} iff 
for every state $\alpha$ on $A$, there exists a model $B$, a composite system $AB$, and a correlating bipartite state $\omega$ on $AB$ such that $\omega_1 = \alpha$. 

The Correlation condition (a dilation principle, like Steering and Iso-dilation) is by no means obvious on purely 
operational grounds. On the other hand, something like it is needed if we are to be able to capture measurement processes ``internally", that is, in terms of the resources available in $\C$. For a further discussion of this point, see \cite{Wilce10}.  

As noted in \cite{Wilce10, Wilce11b}, the correlation condition implies a kind of spectral decomposition for states:

{\bf Lemma 10:} {\em Let $A$ be sharp and satisfy correlation. Then for every state $\alpha$ on $\E$, there exists a test $E \in \A(A)$ and convex coefficients $t_x$ with  $\alpha = \sum_{x \in E} t_x \delta_x$. } 

{\em Proof:} Let $\alpha = \omega_1$ where $\omega$ correlates $E$ with $F$ along $f$. Then, by (\theequation),  \[\alpha \ = \ \sum_{y \in F} \omega_2(y) \omega_{1|x} \ = \ \sum_{x \in E} \omega_2(f(x)) \delta_{x}.\] Set 
$\omega_{2}(f(x)) = t_x$. $\Box$ 

{\bf Definition 26 (Filtering Condition):} $A$ satisfies the {\em filtering condition} iff for every test $E \in \A$, 
and every set of constants $0 < t_x \leq 1$, there exists an affine automorphism $\Phi \in \E_+$ with $\Phi(x) = t_x x$. I'll call such an automorphism 
a {\em filter on $E$}. 

Filtering is a reasonable assumption. If we think of a test $E$ as, e.g., an array of detectors, then the axiom simply asserts that we can independently attenuate the reliabilities of these detectors --- which, in practice, we can certainly do. 

Now suppose that $A$ is sharp, and let $\delta_x$ denote the unique normalized state on $\E$ with $\delta_x(x) = 1$. If $\Phi$ is a filter on $E \in \A(A)$ with $\Phi(x) = t_x x$, $t_x > 0$, then   
\[\Phi^{\ast}(\delta_x)(x) = \delta_x (t_x x) = t_x \delta_x (x) = t_x,\] 
and similarly, $\Phi^{\ast}(\delta_x)(y) = 0$ for $y \perp x$. It follows that $t_{x}^{-1} \Phi^{\ast}(\delta_x) = \delta_x$, 
i.e., $\Phi^{\ast}(\delta_x) = t_x \delta_x$. 

As observed in \cite{Wilce11b}, we now have 

{\bf Lemma 11:} {\em Let $A$ be sharp, state-complete, fully $G$-symmetric, and satisfy both the correlation and filtering axioms. Then $\V(A)_+$ is homogeneous.} 

{\em Proof:} Let $\alpha$ and $\beta$ be normalized states in the interior of $\E_+$. We wish to find some order-automorphism of $\E$ taking $\alpha$ to $\beta$. By Lemma 10, 
we can expand $\alpha$ and $\beta$ as $\alpha = \sum_{x \in E} t_x \delta_x$ and $\beta = \sum_{y \in F} s_y \delta_y$ for some tests $E, F \in \A$. Since $\alpha$ and $\beta$ are interior, $t_x > 0$ and $s_y > 0$ for all $x \in E$ and $y \in F$. Let $f : E \rightarrow F$ be any bijection, 
and let $\Phi$ be a filter on $E$ taking each $x \in E$ to $m_x x$, where $m_x = s_{f(x)}/{t_x}$. Then we have 
\[\Phi^{\ast}(\alpha) = \sum_{x \in E} t_x \Phi^{\ast}(\delta_x) = \sum_{x \in E} t_x m_x \delta_x = \sum_{x \in E} s_{f(x)} \delta_x.\] 
By full symmetry, $f$ extends to a symmetry $g \in G$; applying this, we have 
\[g\Phi^{\ast}(\alpha) = \sum_{x \in E} s_{f(x)} g \delta_x = \sum_{x \in E} s_{f{x}} \delta_{f(x)} = \sum_{y \in F} s_y \delta_y = \beta.  \  \Box\] 
 
Recall that if $(\bar{A},\gamma_A)$ is a {\em strong} conjugate for $A$, then for every state $\alpha \in \Omega$, there exists an equivariant state $\omega^{\alpha}$ on $A \overline{A}$ with $\omega_1 = \alpha$, and correlating some test $E \in \A(A)$ with the conjugate test $\bar{E} \in \A(\bar{A})$ along $x \mapsto \bar{x} := \gamma_{A}(x)$. This gives us the correlation property,  and also, 
if $A$ is sharp and state-complete, self-duality (by Theorem 1).  Thus, we have 

{\bf Theorem 5:} {\em Let $A$ be a sharp, state-complete, irreducible bi-symmetric model having a strong conjugate 
and satisfying the filtering condition. The $\E(A)_+$ is homogeneous and self-dual.} 

We also have %the third of our advertised results, Theorem C, from the Introduction, which I'll reformulate as

{\bf Theorem 6:} {\em Let $\C$ be an image-closed, dagger-monoidal probabilistic theory, in which every 
system is bi-symmetric and state-complete. If $A \in \C$ is sharp and satisfies the Correlation and Filtering conditions, 
then $\E(A)_+$ is homogeneous and self-dual. }

(Notice that here, as in Theorem 4b, image-closure can be dropped, if we are willing to concentrate on incompressible models.)\\

%Notice that in Theorems 1, 2 and 3, I've resisted the obvious urge simply to assume that all systems in $\C$ are %sharp. This is because thre is a tension between sharpness and image-closure: a surjective image of a sharp model %need not be sharp. We could reformulate all of these results by focussing on incompressible models: in any 
%dagger-monoidal category of sharp, 

%{\bf 6.3 Remarks on Quantum Axiomatics} 

%Note here that our strong correlation axiom, plus 2-symmetry plus filtering, leads to spectrality leads to homogeneity, WHICH %LEADS TO IRREDUCIBILITY (??). Thus, no need for image-closure. 

%[Compare with image-closed, dagger-monoidal 2-symmetry; steering; compare with CDP axioms etc. (BUT: keep this 
%section fairly brief!] \\

\section{Conclusion and Speculations}

The foregoing results show that the Jordan structure of finite-dimensional QM emerges very naturally from a few relatively 
simple constraints having reasonably clear operational or physical meanings. Or, better to say, follow from {\em any of 
several different} clusters, or packages, of such constraints. Two of these are given in Theorems 5 and 6. Some others:

\begin{quote}
(1) Individual systems are bi-symmetric, state-closed, irreducible, and has a conjugate system with an iso-correlator. Every interior state can be reversibly prepared from the maximally mixed state.\\

(2) Individual systems are sharp, state-closed, irreducible, fully symmetric, and satisfy both the strong correlation and the filtering condition. \\

(3) Systems collectively form an image-closed category with a dagger-monoidal representation, and individually are sharp, bi-symmetric
and satisfy the steering condition.
\end{quote} 

%%Dagger-monoidal, image-closed cat $\C$ of state-closed, fully-symmetric models, with correlation and filtering. 
 
Obviously, though, there's much left to do. Regarding (3), for example, while existence of a symmetric monoidal structure is not usually viewed as problematic, the existence of a dagger  cries out for further explanation. One would like to find a compelling physical or operational interpretation for such a structure. 
%[Using [BDW], can we show that the existence of correlators gives this structure?] 
(One attractive, though at this point vague, idea is that a dagger corresponds to a global time-reversal symmetry.) 

%A sharper as presently developed, the results obtained here are aesthetically unappealing: they involve too many moving parts. %It is likely, however, 
%that the apparatus can be simplified a good deal --- for example full symmetry and the filtering axiom are, intuitively, %manifestations of the same idea, namely, that a classically reversible process (e.g., a permutation, or a filtering operation), %associated with the outcomes of a single test, should extend to a physically reversible process on the entire system. Also, 
%I have made no special effort to trace the logical interrelations among the various conditions considered here, so there 
%is the possibility that some of the ``packages" considered above are redundant. 

To all of these examples, there is an aesthetic objection: there are too many moving parts. It is likely, however, that the apparatus 
can be simplified. For example, there is a sense in which both full symmetry and filtering are expressions of the same 
idea: that any classically allowed, reversible process acting on the probabilistic apparatus associated with a single test, 
should extend to an irreversible process acting on the entire system. In terms of a slogan: any classically reversible 
process corresponds to a physically reversible process. %[Reversibility discussed above...]  
Finally, it would be very desirable to replace the image-closure condition with some kind of reduction theory, according 
to which all systems in $\C$ simply {\em are} direct sums, in some suitable sense, of irreducible systems. At present, I do 
not see how to obtain such a theory by anything short of {\em fiat}, but this may simply reflect lack of sufficient effort, 
or wit, on my part.\footnote{Alternatively, one could hope to show that (perhaps in the presence of other constraints), homogeneity already implies irreducibility. This is true, for example, if the group $G$ comprises all unit-preserving order-automorphisms in the connected component of the identity of $\Aut(\E)$.}

I have made no real effort to establish in detail how the various conditions enumerated here depend on one another, so there is the possibility that, given some of them, others are simply redundant. It is also perfectly conceivable that these conditions  are stronger than necessary. For example, I haven't checked to see whether every simple Jordan model has a 
conjugate, or satisfies filtering. At a more fundental level, it remains an important open question whether there exist any {\em non}-$C^{\ast}$-algebaic dagger-symmetric monoidal categories of formally real Jordan algebras.  

I want to emphasize again that local tomography has played no role here.  In a forthcoming paper \cite{BWta} with Howard Barnum, it will be shown that if $\cal E$ is a dagger-HSD category of order-unit spaces with non-signaling, locally tomographic composites, and if $\cal E$ contains a model having the structure of a qubit, then it is a category of finite-dimensional complex matrix algebras. 

{\bf Acknowledgements} I am indebted to David Feldman, Howard Barnum and Jochen Rau for helpful comments and questions regarding earlier versions of this paper.

%{\bf References} 

\end{document}